\documentclass[lettersize,journal]{IEEEtran}
\usepackage{amsmath,amsfonts}
\usepackage{algorithmic}
\usepackage{algorithm}
\usepackage{array}
\usepackage{amssymb}
\usepackage[caption=false,font=normalsize,labelfont=sf,textfont=sf]{subfig}
\usepackage{textcomp}
\usepackage{stfloats}
\usepackage{url}
\usepackage{verbatim}
\usepackage{graphicx}
\usepackage{cite}
\usepackage{bm}

\begin{document}

\title{Pilot-Free Unsourced Random Access Via Dictionary Learning and Error-Correcting Codes}
%\author{Zhentian Zhang

\author{Zhentian Zhang,~\IEEEmembership{Student Member,~IEEE,} Jian Dang,~\IEEEmembership{Senior Member,~IEEE,}
Zaichen Zhang,~\IEEEmembership{Senior Member,~IEEE,}
Liang Wu,~\IEEEmembership{Senior Member,~IEEE,}
Bingcheng Zhu,~\IEEEmembership{Member,~IEEE,} 
and
Lei Wang
        % <-this % stops a space
\thanks{ }% <-this % stops a space
\thanks{Zhentian Zhang, J. Dang, Zaichen Zhang, L. Wu, B. Zhu and L. Wang are with the National Mobile Communications Research Laboratory, Frontiers Science Center for Mobile Information Communication and Security, Southeast University, Nanjing, 210096, China. J. Dang, Zaichen Zhang, L. Wu, B. Zhu and L. Wang are also with the Purple Mountain Laboratory, Nanjing 211111, China (e-mail: zhangzhentian@seu.edu.cn; dangjian@seu.edu.cn; zczhang@seu.edu.cn; wuliang@seu.edu.cn; zbc@seu.edu.cn; wang\_lei\_seu@seu.edu.cn).}% <-this % stops a space
\thanks{Corresponding author: J. Dang (dangjian@seu.edu.cn)}}% <-this % stops a space



\maketitle

\begin{abstract}
Massive machine-type communications (mMTC) or massive access is a critical scenario in the fifth generation (5G) and the future cellular network. With the surging density of devices from millions to billions, unique pilot allocation becomes inapplicable in the user ID-incorporated grant-free random access protocol. Unsourced random access (URA) manifests itself by focusing only on unwrapping the received signals via a common codebook. In this paper, we propose a URA protocol for a massive access cellular system equipped with multiple antennas at the base station. The proposed scheme encompasses a codebook enabling construction of sparse transmission frame, a receiver equipped with dictionary learning and error-correcting codes and a collision resolution strategy for the collided codeword. Discrepant to the existing schemes with necessary overhead for preamble signals, no overhead or pre-defined pilot sequences are needed in the proposed scheme, which is favorable for energy-efficient transmission and latency reduction. Numerical results verify the viability of the proposed scheme in practical massive access scenario. 
\end{abstract}


\section{Introduction}
\IEEEPARstart{M}{assive} machine-type communications (mMTC), also known as massive connectivity or massive access, aims at ensuring efficient, robust and ubiquitous wireless communications for billions of energy-constrained devices\cite{ref1}. A typical application of mMTC is the Internet of Things (IoT). Different from the traditional communications mainly designed for human-type interaction, such as Long Term Evolution (LTE), there are several notable features in mMTC scenario. For example, 1) Uplink-driven sporadic traffic: only a fraction of devices are active at a certain transmission instant and others retain silent 2) Short-packet transmission: normally, only a small volume of bits is transmitted by a device and for the concern of resource efficiency, short-packet transmission is more preferable 3) Energy-efficient communication protocol: to prolong the life span of battery, ingenious strategy has to be employed to reduce the power consumption. Thus, under the background of mMTC, new transmission protocols are in demand to support the massive connectivity.

In a cellular system, such as LTE, a grant-based random access protocol has been widely adopted, in which an active device has to send request and obtain permission from the base station (BS) to access the network. However, with the density of active users strikingly ascending, lack of orthogonal preamble sequences and the consequent high level of collision give rise to high latency or access failure between devices and BS\cite{ref2,ref3}. To mitigate the problems above, grant-free random access protocol has been identified as an enabler for mMTC\cite{ref4}. In grant-free access, active devices transmit signals without preceding requesting and approval from the BS. Specifically, active devices send information appended with unique non-orthogonal preamble sequences directly without getting confirmation from the BS. Thus, the access latency and the transmission overhead are significantly decreased. In general, the receiver aims to conduct active user detection from all codewords and achieve information restoration. Owing to the sporadic traffic, those problems can be formulated into a compressed sensing (CS)-based sparse recovery framework\cite{ref5} or a covariance-based problem\cite{ref6}. However, due to the non-orthogonality, the transmitted preamble signals suffer from severe co-channel interference, which raise higher demands on the activity detection algorithms. Compared with the grant-based protocol, the access latency in grant-free paradigm is reduced at the cost of higher computational complexity. Nonetheless, as the number of device surges, larger codebook needs to be designed to support preamble allocation. However, the continuing codebook extension eventually produces unbearable computational complexity. 

Recently, another kind of grant-free random access called unsourced random access (URA) has been proposed and received great interests\cite{ref7}. In URA, instead of being assigned with preambles individually, active devices share a common codebook and if the devices wish to send signals for identification or authorizations, they can add these into frames as extra payloads. Devices send signals directly to the BS and the task at the BS is to restore the information without prescheduling and identifying devices, leading to the so-called unsourced property, i.e, the unsourced random access casts off the need of coordination center\cite{ref8}. Having been proven by Polyanskiy, URA can robustly support substantial active devices in manner of per-user error probability. Practically, how to provide desirable reliability for single user remains an open discuss. How to avoid severe interferences between users in uplink transmission and how to design practical coding structure are two key factors for URA. 
Interferences mitigation among active devices and resource multiplexing directly influence the total volume of the system.
SCMA\cite{ref131} is a classic resource multiplexing enhancement for physical layer in time domain. In SCMA, the transmission patterns are arranged sparsely by allocating codebook individually and then the system capacity is enhanced. However, this is inapplicable in URA for the concern of the shared common codebook. Furthermore, after receiving the transmitted signals, detection and estimation tasks need to be carried out to restore information. Normally, preambles are used as extra payload to achieve estimation tasks, whereas tolerance for extra payload are limited for the short-packet transmission concern in mMTC. The structure for these tasks correlates with the coding design in URA. A novel solution with practical codebook and feasible coding scheme with less payload is demanded to harness URA to enhance the massive connectivity. In this paper, we consider an uplink transmission in the URA scenario with multiple input multiple output (MIMO) where a massive number of single antenna devices access the BS with multiple antennas.

%Specifically, a URA protocol equipped with a sparse transmission frame depending on codebook and a receiver processing combined with dictionary learning and error correcting codes is performed to support massive connectivity without any pilot sequences or pilot-like bits.

\subsection{Related Works}
From the universal view of uplink transmission, the transmission structure design is crucial for enhancing the overall transmission capacity. Specifically, direct spreading transmission and its intermediate variants\cite{ref12,ref13} are widely used. Direct spreading can be modeled as the Gaussian multiple access channel (GMAC) of $K_{a}$ users with $k$-bit input and $n$-length overlapped codeword output. For T-Fold ALOHA\cite{ref12}, “T-Fold” means the total amount of users collided on any slots is constrained under a threshold $T$. It means the receiver only needs to have the processing capability to support certain amount of user in one slot. T-Fold repetition slotted ALOHA (T-Fold RSA) \cite{ref13} designates a subset of users or all users to repeat packets for certain or random times. This can generate diversity gain by packet coding and enable packet-level successive interference cancellation at the receiver. On this ground, sparsity in transmission pattern is a key factor for capacity enhancement.

The other line of work in URA is coding scheme design incorporating the work of activity detection and information restoration. A coded compressed sensing (CCS) scheme\cite{ref14} adopts the divide-and-conquer strategy for detection and restoration. The frame to be transmitted is divided into small segments which are transmitted in sequential order. And redundant parity bits are added to stitch the segments together afterwards. This segmentation reduces the need of larger sensing matrix for compressed sensing (CS) procedure. Small sensing matrix leads to a relaxation on computational complexity at the receiver. Yet, the parity bits aiming at stitching all segments deplete the transmission efficiency. Different to CCS, interleave division multiple-access (IDMA)\cite{ref108} only divides message frame into two segments. The first segment enables activity detection from codebook and the detected codewords also correspond to the interleaving patterns in second segment. In other words, the interleaving pattern of the second part of the frame is embedded into the first segment. Besides, interleaving is conducted after zero-padding in the second segment. Through zero-padding and interleaving, the interferences between users is depleted. Thus, the transmission capacity is enhanced. \cite{ref106,ref125} design their structure with different channel codings and all consider GMAC with perfect channel state information (CSI).  \cite{ref106} adopts the structure of low density parity check (LDPC) code and \cite{ref125} selects polar code as the basic multiple access code. However, perfect CSI cannot be an established assumption under MIMO scenario. In \cite{ref123}, estimation of MIMO channel is done by utilizing dictionary learning (DL)-based sensing matrix update method. \cite{ref122} allocates user-discrepant pilot-like priors to assist the adoption of DL method, especially the ambiguity problems encountered after decomposition of the observations which will be explained in Section II. This scheme achieves good performances in user-ID based grant-free MIMO system. However, its use of giant codebook to achieve activity detection lacks efficiency and due to the limited priors in URA, similar approaches need further researches.

 %designs a coding scheme by the spirit of IDMA and choosing the structure of LDPC. The header performs the task of CSI estimation under MIMO scenario and delivers the parameter of interleaving pattern. This interleaving pattern and the estimated CSI are than combined for the latter data restoration. 

%This paper takes the strategy of direct spreading. The bits to be transmitted are arranged into a sparse frame. The sparse permutation pattern of the nonezero elememts in the frame is determined by the chosen codeword from the common codebook. The codeword's sparsity feature is directly correlated with the activity detection and data restoration at the receiver.  

\subsection{Contributions and Organization}
This paper introduces a URA protocol for massive access in a MIMO cellular system where single antenna devices interact with a base station with multiple antenna. The proposed scheme consists of a sparse transmission pattern-oriented codebook, a DL and error-correcting codes (ECC)-based receiver and a collision resolution procedure. The active devices initially choose codewords from the common codebook and then generate a sparse frame into which the modulated symbols are spread in accordance with the selected codewords. The symbols' sparsity pattern is directly correlated with the latter activity detection and information restoration. For the receiver, the BS restores the sparse frame and modifies the estimated channel matrix by DL and ECC. The sparsity in the frame promotes the realization of DL. The combination with DL and ECC achieves the proposed pilot-free URA protocol. Especially, discrepant to the existing work of arts, the need of preambles signal for various uses, such as channel estimation, activity detection and so on, is essentially not required, which is favorable in terms of short-packet transmission for power-limited devices in mMTC.

With the proposed DL-ECC-based protocol, a BS with multiple antennas can support a massive amount of devices without pre-coordination in URA manner. And a collision resolution procedure designed by feature of DL and ECC is explained. Numerical results illustrate this proposed protocol's viability and compare its performance with other DL-based method. The organization of this paper is as following: Section II describes the system model and gives a brief background introduction on DL and ECC. The ambiguity problems in DL are also elaborated. In Section III, the proposed receiver structure is elaborated and, Section IV illustrates the numerical results of system performances. Finally, Section V draws the conclusions.

\textit{Notations}: For a matrix $\mathbf{A}$, $\mathbf{A}^{-1}$ denotes the inverse if $\mathbf{A}$ is invertible, $\mathbf{A}^{*}$ and $\mathbf{A}^{\text{T}}$ denote the complex conjugate and the transpose of $\mathbf{A}$, and $\left [ \mathbf{A} \right ]_{i,j}$ represents the element at the $i\text{-th}$ row and $j\text{-th}$ column of the matrix $\mathbf{A}$. $\left [ \mathbf{A} \right ]_{i,:}$ means the $i$-th row of the matrix and $\left [ \mathbf{A} \right ]_{:,j}$ is the $j$-th column. $\textbf{zeros}(n,m)$ creats a all zero matrix with $n\times m$ size. For a vector, $\mathbf{a}$ denotes a column vector and $\mathbf{a}^{\text{T}}$, a row vector.
Semicolon is used to isolate row vectors in a matrix $\mathbf{A}=\left [ \mathbf{a}_{1}^{\text{T}}; \mathbf{a}_{2}^{\text{T}};\dots ; \mathbf{a}_{n}^{\text{T}} \right ] $ and comma is used to isolate column vectors $\mathbf{A}=\left [ \mathbf{a}_{1},\mathbf{a}_{2},\dots , \mathbf{a}_{n} \right ]$.
$\left [\mathbf{a} \right ]_{i}$ means the $i\text{-th}$ element. $\left \| \mathbf{a} \right \|_{p}$ stands for $p$-norm, where $\left \| \mathbf{a} \right \|_{0}$ means the number of none-zero elements in the vector. $diag(\mathbf{a})$ is a square matrix whose diagonal elements are $\mathbf{a}$. For a set $A$, $\left | A  \right |$ is the cardinality of $A$ and $\left \| A \right \|_{0}$ means the number of nonezero elements in the set. $(A_{1}-a)$ means a new set by popping out the element $a$. For a scalar $a$, $\left | a  \right |$ is the absolute value of $a$. $\mathcal{CN}\left ( \bm{\mu} ,\bm{\Sigma} \right )$ represents the multivariate complex Gaussian distribution with mean $\bm{\mu}$ and covariance matrix $\bm{\Sigma} $. 
\section{System model}
\subsection{Uplink Transmission Model}
\begin{figure}[!t]
  \centering
  \includegraphics[width=3.4in]{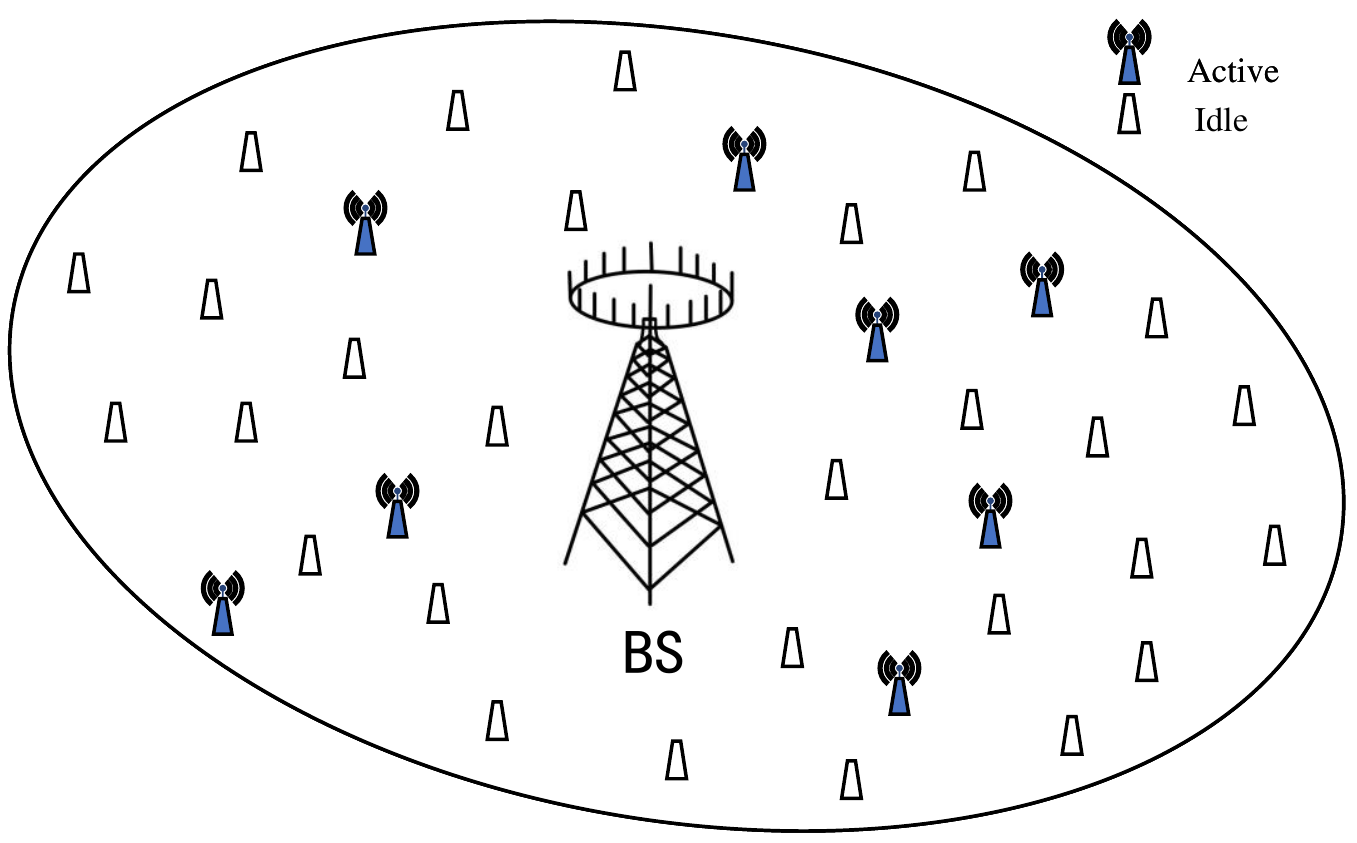}
  \caption{Illustration of sporadic uplink transmission in mMTC system with a MIMO base station.}
  \label{SystemModel}
  \end{figure}
  As showed in Fig. 1, consider the uplink of a single-cell cellular network consisting of $K_{tot}$ single-antenna devices, which are being served by a BS equipped with $M$ antennas. Sporadic activity is assumed, i.e., $\mathcal{A} \subset \left \{ 1,2,...,K_{tot} \right \} $ denotes a set of active users within a coherence time and the set cardinality is $\left|\mathcal{A}\right|=K_{a} \ll K_{tot}$. Each device has $B$ bits of information to be coded and embedded into a frame transmitted in each of $L$ channel uses. $\mathbf{x}_{k}\in \left \{ 0,1 \right \} ^{B\times 1}$ denotes $k$-th active device's binary message. $f\left ( \cdot  \right )$ is an encoding function in the proposed scheme, representing error-correcting codes (ECC) encoding and selecting codeword from common codebook and symbol modulation. This encoding procedure will be elaborated in Section III. By assuming the synchronized transmission among devices, the received signal can be represented as 
\begin{equation}
\label{system1}
\mathbf{Y}= \sum_{u\in\mathcal{\mathcal{A} }} \sqrt{\rho_{u}} \mathbf{h}_{u}f(\mathbf{x}_{u}^{\text{T}})+\mathbf{N}
\end{equation}
where $\rho_{u}$ is the received power per-symbol, $\mathbf{h}_{u}\in \mathbb{C}^{M\times 1}$ is the $u$-th active device's channel vector and $\mathbf{h}_{u}\sim \mathcal{CN}\left ( \mathrm {0},\mathbf {I} \right )$ the Rayleigh fading model is considered, $\mathbf{x}_{u}^{\text{T}}\in \left \{ 0,1 \right \} ^{1\times B }$ is the binary messages row vector and $f(\mathbf{x}_{u}^{\text{T}})\in \mathbb{C}^{1\times L}$ is the $\mathbf{x}_{u}^{\text{T}}$ output from the encoder, denoting the codeword of the active device $u\in \mathcal{A}$, $\mathbf{N}\in \mathbb{C}^{M\times L}$ is the additive white Gaussian noise matrix whose elements are independently distributed as $\mathcal{CN}\left ( 0,1 \right )$.   

\subsection{Dictionary Learning}
In this section, the relationship between sparse representation of signals and dictionary learning is briefly introduced. Typical recovery from noisy measurements \cite{ref110} can be formulated into the estimation of $\mathbf{\tilde{x} }\in\mathbb{C}^{n\times1}$ from
\begin{equation}
\label{CS1}
\mathbf{y}=\mathbf{D \tilde{x}+  n}
\end{equation}
where $\mathbf{y}\in\mathbb{C}^{m\times1}$ is a measurement vector and $\mathbf{D}\in\mathbb{C}^{m\times n}$ is a known sensing matrix or \textit{dictionary} matrix, and $\mathbf{n}$ is the noise vector. In CS, the sensing matrix is often a fat matrix, i.e. $m \ll n$, it becomes an under-determined scenario that a unique solution hardly exits, and infinite solutions are often the case. However, by exploiting the sparsity feature of $\mathbf{\tilde{x} }$, the sparsest solution can be found\cite{ref109}. By the sparse attribute of $\mathbf{\tilde{x} }$, this NP-hard, yet soluble problem can be well-settled by various methods. For example, in \cite{ref116}, orthogonal matching pursuit (OMP) algorithm is a greedy and easily implemented algorithm that can recover and also control the sparsity of $\mathbf{\tilde{x} }$. Nevertheless, other solutions \cite{ref111,ref112,ref113,ref114,ref115} have been well discussed, amenable to achieving various goals by exploiting sparsity feature in mMTC, such as activity detection and channel estimation.

While CS uses linear measurements and under-determined sensing matrix to carry out sparse recovery, dictionary learning (DL) generates dictionary or sensing matrix from received signals.
Generally, dictionary learning can be deemed as the factorization of $\mathbf{Y=\tilde{D} \tilde{X}+N}$, where $\mathbf{N}$ is the noise matrix, $\mathbf{Y}=\left [ \mathbf{y}_{1}, \mathbf{y}_{2},..., \mathbf{y}_{L} \right ] $, $\mathbf{X}=\left [ \mathbf{\tilde{x}}_{1}, \mathbf{\tilde{x}}_{2},..., \mathbf{\tilde{x}}_{L} \right ] $ . The very basic idea is to decompose observation $\mathbf{Y}$ with a prespecified or primitive dictionary $\mathbf{\tilde{D}}$ to produce coefficients, i.e., the observation is approximated to the linear combination of dictionary's column vectors. Then, by updating dictionary or coefficients or both, the deviation is further reduced. For examples, initially, OMP is taken to decompose $\mathbf{Y}$ into the product of a primitive dictionary $\mathbf{\tilde{D}}$ and $\mathbf{\tilde{X}}$. Subsequently, method of optimal direction (MOD) algorithm\cite{ref120} modifies dictionary to make the approximation more accurate with fixed coefficients or K-SVD\cite{ref121} algorithm iteratively updates both the dictionary and coefficients together in column-wise order, via singular value decomposition. Specifically, the above can be formulated as 
\begin{equation}
  \label{DL1}
  \begin{split}
  &\mathop{\arg\min}\limits_{\mathbf{\tilde{D}}, \left \{\mathbf{\tilde{x}}_{i}\right \}}\sum_{i=1}^{L} \left \| \mathbf{\tilde{x}}_{i} \right \|_{0}, \\
  &\text{s.t.}\quad \sum_{i=1}^{L} \left \| \mathbf{y}_{i}-\mathbf{\tilde{D}}\mathbf{\tilde{x}}_{i}  \right \|_{2} \le \epsilon
  \end{split}
  \end{equation}
where $\epsilon$ is the acceptable error, with which the equation yields the $\mathbf{\tilde{X}}$ with desired sparsity. Equivalently, another formulation is
\begin{equation}
  \label{DL2}
  \begin{split}
  &\mathop{\arg\min}\limits_{\mathbf{\tilde{D}}, \left \{\mathbf{\tilde{x}}_{i}\right \}} \sum_{i=1}^{L} \left \| \mathbf{y}_{i}-\mathbf{\tilde{D}} \mathbf{\tilde{x}}_{i}  \right \|_{2},\\ 
  &\text{s.t.}\quad \sum_{i=1}^{L} \left \| \mathbf{\tilde{x}}_{i} \right \|_{0} \le C_{0}
  \end{split}
  \end{equation}
which sets a constant $C_{0}$ to constrain the sparsity of $\mathbf{\tilde{X}}$.
Notably, There are two inherent ambiguities in DL, supposing $\mathbf{D}$ and $\mathbf{X}$ are the original matrix. 1) Permutation ambiguity: it implies that the rows and columns of the solution $\mathbf{\tilde{D}}$ and $\mathbf{\tilde{X}}$ permute at random, i.e., $\mathbf{D X}=\tilde{\mathbf{D}}\mathbf{\Pi}  \times  \mathbf{\Pi}^{\text{T}}    \tilde{\mathbf{X}}$, $\mathbf{\Pi}$ is a square permutation matrix with only one element valued 1 in each column and row. $\tilde{\mathbf{D}}\mathbf{\Pi}$ is the column-permuted version of $\mathbf{D}$ and $\mathbf{\Pi}^{\text{T}} \tilde{\mathbf{X}}$ is the row-permuted version of $\mathbf{X}$. 2) Scalar ambiguity: it means when the dictionary matrix and coefficients matrix are multiplied with a constant and its inverse, the product is identical to the original, i.e., $\mathbf{DX}=\tilde{\mathbf{D}}\mathbf{\Lambda}^{-1} \times \mathbf{\Lambda} \tilde{\mathbf{X}}$, $\mathbf{\Lambda}$ is a diagonal constant matrix. These two are inherent problems in DL and need to be cautiously dealt with when the permutation and the scalar of the solution are crucial. In Section III, how to tackle with these ambiguities in the proposed scheme will be elaborated.
In the proposed scheme of this paper, how to mingle DL and ECC together to achieve activity detection and information restoration at the BS is first proposed and will be further discussed in Section III.
%\begin{figure}[!t]
  %\centering
  %\includegraphics[width=3.4in]{ECCProcess.pdf}
  %\caption{The paradigm with error-correcting coding before transmitting and after receiving the signals.}
  %\label{ECCProcess}
  %\end{figure}

\section{DL-ECC-based receiver design}
In this section, a practical UMA scheme consisting of DL and ECC is introduced in accordance with the model in (\ref{system1}). Initially, a sparse frame construction is elaborated to insert symbols into the frame. Next, a receiver processing scheme, incorporating both DL and ECC, is designed to fulfill activity detection and information recovery from the received signals. Eventually, to conquer the low potential codeword collision in URA, a collision resolution is designed by the feature of ECC activity detection.

\subsection{Frame Structure and Sparsity Construction}

\begin{figure}[!t]
  \centering
  \includegraphics[width=3.5in]{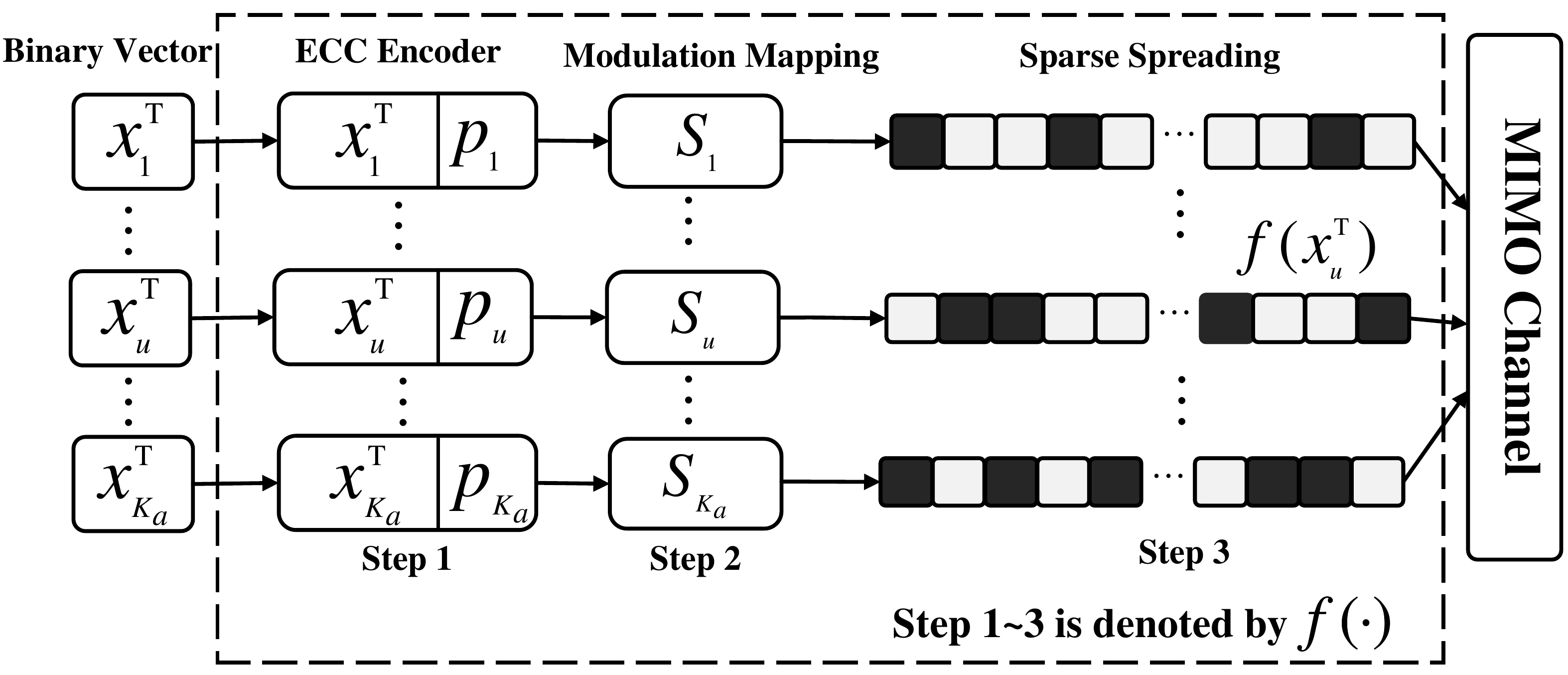}
  \caption{All coding procedures of the proposed scheme before transmission are illustrated, incorporating ECC encoding, modulation mapping and codeword-oriented sparse spreading.}
  \label{BeforeTransmission}
\end{figure}

This subsection elaborates the compositions of the frame and the transmission strategy. The sparse frame structure enables active codeword identification and information restoration via the dictionary learning and ECC. Supposing an active user $u \in \mathcal{A}$ is to transmit a binary vector $\mathbf{x}_{u}^{\text{T}}$ within a frame of $L$ channel uses. Fig. 2 illustrates the whole procedures before transmission. The vector $\mathbf{x}_{u}^{\text{T}}$ is coded into $f(\mathbf{x}_{u}^{\text{T}})$ by the following procedures, namely ECC encoding, modulation mapping and sparse spreading by codeword pattern. Without loss of generality, Low Density Parity Check (LDPC) code represents the ECC encoding for the following description. For the ease of description, $f\left ( \cdot  \right )$ represents the aggregation of three maneuvers the binary vector undergoes. 

Initially, LDPC encoding is conducted. The parity check binary vector $\mathbf{p}_{u}$ with $p$ bits is constructed in accordance with the prescribed parity check matrix $\mathbf{H}$ which can be modified to satisfy different bit rate. Once the LDPC encoding stage finishes, the parity check bits are appended to the rear part of the preliminary messages. Next, vector $\left [ \mathbf{x}_{u}^{\text{T}},\mathbf{p}_{u} \right ] $ with $(B+p)$ bits is mapped into complex-valued symbol vector $\mathbf{S}_{u}$ by modulation method. In this paper, Quadrature Phase Shift Keying (QPSK) is considered. Lastly, the frame is constructed by spreading symbols in accordance with the sparsity pattern from the selected codeword. In our schemes, codewords the active users choose also control the sparsity pattern of the transmission. However, discrepant to the work of art, the proposed scheme casts the prior setting of pilot sequence or pilot-like bits, and to follow the nature of the UMA, no codeword is uniquely assigned to users. $\mathbf{c}_{u}^{\text{T}}\in \left \{ 0,1 \right \} ^{1\times L} $ is the $u$-th active user's codeword. It contains $S=\frac{B+p}{2} $ nonezero elements valued one and $L-S$ zeros. $S$ equals the number of the modulated symbols and $L$ equals the length of the frame. Besides, $S\ll L$, i.e., very few elements in $\mathbf{c}_{u}^{\text{T}}$ are nonezero and $L/S$ quantifies how sparse the vector is. The zeros denote the silent slots which means no signal is sent, and the symbols awaiting transmission are scattered on the position of nonezero elements. This sparse spreading procedure is illustrated in Fig. 2-Step 3, where dark compartment is the spread symbol and the shallow areas are silent slots. For example, symbol vector $\left [ S_{1},S_{2},...,S_{k} \right ]$ is inserted into frame $\left [ S_{1},0,0,S_{2},0,...,0,0,S_{k},0 \right ]$ by the sparsity pattern of the selected codeword $\left [ 1,0,0,1,0,...,0,0,1,0 \right ]$. When the slot is a silent one, no signals are transmitted. This sparsity of frame is later utilized to achieve activity detection and information restoration.

%\begin{figure}[!t]
 % \centering
 % \includegraphics[width=3.5in]{SparseSpreading.pdf}
  %\caption{Illustration for sparse spreading to construct a sparse frame by the sparsity pattern of codeword. Binary messages are firstly modulated into symbols $S_{i}$. $M$ bits are converted into one symbol according to modulation order. Then, symbols, $S_{i}$, are sparsely spread by codeword to construct the frame. And the shallow-colored compartment represents the idle slot where no symbols are transmitted at all. Symbol vector $\left [ S_{1},S_{2},...,S_{k} \right ]$ is inserted into frame $\left [ S_{1},0,0,S_{2},0,...,0,0,S_{k},0 \right ]$ by the sparsity pattern of the selected codeword $\left [ 1,0,0,1,0,...,0,0,1,0 \right ]$.}
  %\label{SparseSpreading}
%\end{figure}

Note that the total amount of permutations of nonezero elements in a single codeword is obedient to $\binom{L}{S}$, where the $\binom{L}{S}$ represents the combination number formula, which signifies a tremendous amount of various codewords and reveals a rather low likelihood that the collision between active users would happen when codewords are chosen at random, meanwhile, $L\gg S$ makes the sparsity of transmission vectors feasibly controllable and thus conducive to the later DL procedure.

\subsection{Joint Active Device And information Detection}
In this subsection, a receiver designed to jointly perform active device and information detection is described. The following covers the content on how to correlate UMA with DL, and how to solve the ambiguity problems in DL by ECC. 

\textit{1) DL problem formulation:} During a frame of time, (\ref{system1}) can be rewritten as
\begin{equation}
\label{DLtoCS}
\begin{split}
  \mathbf{Y}=\mathbf{G X}+\mathbf{N}
\end{split}
\end{equation}
where $\mathbf{G}=\left[\sqrt{\rho_{1}} \mathbf{h}_{1},\sqrt{\rho_{2}} \mathbf{h}_{2},\dots ,\sqrt{\rho_{K_{tot}}}\mathbf{h}_{K_{tot}} \right]\in\mathbb{C}^{M\times K_{tot}}$ is the channel matrix and $M$ is the number Of antenna. Row-wisly, $\mathbf{X}=\left [ f(\mathbf{x}_{1}^{\text{T}});f(\mathbf{x}_{2}^{\text{T}});\dots;f(\mathbf{x}_{K_{tot}}^{\text{T}}) \right ]\in\mathbb{C} ^{K_{tot}\times L}$, $\mathbf{X}$ is a row-sparse matrix. As referred in Section III-A, the signals are transmitted by following a specific sparse pattern determined by the selected codeword, i.e., the row vectors of $\mathbf{X}$ complies with a highly sparse pattern owing to the codeword's feature of $S/L\ll 1$. This also makes $\mathbf{X}=\left [ \mathbf{x}_{1},\mathbf{x}_{2},\dots , \mathbf{x}_{L}\right ]^{^{K_{tot}\times L}}$ a column-sparse matrix. The nonezero elements in $\mathbf{x}_{l}, l\subset \left \{ 1,2,\cdots ,L \right \}$ abides by the Binomial distribution. Specifically, looking at the $l$-th channel use, the received signals can be written as: 
\begin{equation}
\label{l_channel_use}
\mathbf{y}_{l}=\mathbf{G x}_{l}+ \mathbf{n}
\end{equation}
where $l\in \left [ 1,L \right ] $, and $\mathbf{y}_{l}$ is the $l$th column of $\mathbf{Y}$. $\mathbf{y}_{l}$ is represented sparsely as the linear combination of the columns in $\mathbf{G}$ and $\mathbf{x}_{l}$ is the coefficient vector. Recalling (\ref{DL2}) and (\ref{l_channel_use}), (\ref{DLtoCS}) can be formulated into a DL problem referred in Section II-B:
\begin{equation}
\label{l_channel_use_DL}
\begin{split}
(\tilde{\mathbf{G}},\tilde{\mathbf{X}})=&\mathop{\arg\min}\limits_{(\mathbf{G, X})} \sum_{l=1}^{L} \left \| \mathbf{y}_{l}-\mathbf{G x}_{l}  \right \|_{2},\\ 
&\text{s.t.}\quad \sum_{l=1}^{L} \left \| \mathbf{x}_{l} \right \|_{0} \le K
\end{split}
\end{equation}

However, direct application of DL algorithms cannot solve the problems in this URA system. The thing is that information lays in the sparse pattern of rows In $\mathbf{X}$. Yet, rows are arbitrarily permuted after DL decomposition. It needs to be further detected to ensure which row is active and to match the row with its codeword to extract desired information, i.e., ambiguity problems needs to be settled, leading to the following tasks.  

\begin{algorithm}[H]
  \caption{Sparse frame-based active codeword detection }\label{alg:alg1}
  \begin{algorithmic}[1]
  \STATE {\textsc{INPUT}} $\bm{\Omega} $, $ \mathbf{C}$, $ \mathcal{\tilde{A}}=\textbf{zeros}(1,1:K_{tot})$
  \STATE \hspace{0.5cm}$ \textbf{Inner Product} \ \mathbf{P=\Omega \cdot C}$
  \STATE \hspace{0.5cm}$\textbf{For} \ i=1:K_{a} \ \textbf{do}$
  \STATE \hspace{1cm}$ \left ( k,n \right )= \textbf{find}( \ \mathbf{P} =max[\mathbf{P}] \ ) $
  \STATE \hspace{1cm}$ \left [ \mathcal{\tilde{A} } \right ] _{k}=n $
  \STATE \hspace{1cm}$ \left [ \mathbf{P}\right ]_{k,:} =\textbf{zeros}(1,K_{tot}) $
  \STATE \hspace{0.5cm}\textbf{end for}
  \STATE {\textsc{OUTPUT}} $ \mathcal{\tilde{A}} $
  \end{algorithmic}
  \label{alg1}
  \end{algorithm}  
\textit{2) Active codeword detection:} Assuming the successful decomposition of the received signals $\mathbf{Y}\approx \tilde{\mathbf{G}} \tilde{\mathbf{X}}$, compared with the desired $\mathbf{X}$, $\tilde{\mathbf{X}}$ can be deemed to be a row-permuted and a scalar-multiplied version due to the permutation and scalar ambiguities. However, the row-wise permutation has no impact on the messages' sequential order. As referred in Section III-A, all active devices randomly choose a sparse codeword from the common codebook and the symbols are transmitted by the order of nonezero elements of the selected codeword, which means $\tilde{\mathbf{X}}$ inherits the sparsity feature of the codeword. Since the column-wise order and the sparsity pattern of $\mathbf{X}$ are well-preserved in $\tilde{\mathbf{X}}$, we can use the sparsity pattern to detect and distinguish the potential active devices. Specifically, after decomposition, we now suppose $\tilde{\mathbf{x}}_{u}^{\text{T}}$ is restored perfectly. It indicates the $\tilde{\mathbf{x}}_{u}^{\text{T}}$'s nonezero elements only exist at the corresponding nonezero elements' position of the selected codeword, i.e., $\tilde{\mathbf{x}}_{u}^{\text{T}}$ and $\mathbf{c}_{u}$ incorporate the same sparsity pattern. Furthermore, owing to $S/L\ll 1$, the inner product between $\mathbf{x}_{u}^{\text{T}}$ and codeword is $\mathbf{c}_{u}^{\text{T}}\cdot \tilde{\mathbf{x}}_{u}=S\cdot a$ and $\mathbf{c}_{\tilde{u} }^{\text{T}}\cdot \tilde{\mathbf{x}}_{u}\ll S\cdot  a$, where $a$ is a complex constant, and $\tilde{u} \in (\mathcal{A}-u)$. This means $\mathbf{c}_{u}^{\text{T}}$ is the most compatible codeword to $\tilde{\mathbf{x}}_{u}^{\text{T}}$ rather than others. Thus, the sparse structure of the frame can be leveraged to achieve active codeword detection and make preparations for the following procedures. 

  To detect every potential active codeword is to pair every rows of $\tilde{\mathbf{X}}\in \mathbb{\mathbb{C}} ^{K_{tot}\times L }$ with all columns of codebook, $\mathbf{C}=\left [ \mathbf{c}_{1},\mathbf{c}_{2},\dots ,\mathbf{c}_{K_{tot}}\right ]\in\left \{ 0,1 \right \} ^{L\times K_{tot}}$, which is equivalent to make inner product between each row of $\tilde{\mathbf{X}}$ and each column of $\mathbf{C}$. First, we extract the sparsity pattern of $\tilde{\mathbf{X}}$ by defining:  
\begin{equation}
  \label{ExtractSparsePattern}
  \left[\mathbf{\Omega}\right]_{i,j} =\left\{\begin{matrix} 
    1,\tilde{\mathbf{X}}_{i,j}\ne 0 \\  
    0,\tilde{\mathbf{X}}_{i.j}=0
  \end{matrix}\right. 
  ,1\le i  \le K_{tot},1\le j  \le L
\end{equation}
and by this way, $\bm{\Omega} =\left [ \bm{\Omega}_{1}^{\text{T}};\bm{\Omega}_{2}^{\text{T}};\dots ;\bm{\Omega}_{K_{tot}}^{\text{T}}\right ]\in \left \{ 0,1 \right \} ^{K_{tot}\times L}$ and have identical sparsity pattern to $\tilde{\mathbf{X}}$. The row vectors $\Omega_{i}^{\text{T}}$ and $\mathbf{x}_{i}^{\text{T}}$ have the exact amount and sequential order of nonezero elements. We define $\mathbf{P=\Omega\cdot C}=\left [ \mathbf{p}_{1}^{\text{T}};\mathbf{p}_{2}^{\text{T}};\dots ;\mathbf{p}_{K_{tot}}^{\text{T}} \right ] \in \mathbb{R} ^{K_{tot}\times K_{tot}}$, where row vector $\mathbf{p}_{i}^{\text{T}}$ denotes the inner product between $\Omega_{i}^{\text{T}}$ and all codewords. Optimally, $\bm{\Omega}_{u}^{\text{T}}\cdot \mathbf{c}_{u}=S$ and $\bm{\Omega}_{u}^{\text{T}}\cdot \mathbf{c}_{\tilde{u}}\ll S$, indicating sparsity pattern matching. However, due to the outer disruptions, the decomposition may not be as smooth as expected, i.e., when $\mathbf{Y}$ is decomposed, the nonezero elements' amount and sequential order of the rows in $\tilde{\mathbf{X}}$, become relatively uncertain because of the noise, and the inner product may not reveal itself as strong as anticipated. To tackle this, we deem the largest element in $\mathbf{P}$ to be the indicator of the potential active codeword. 

The above is described in Algorithm 1, where the output $\bm {\mathcal{\tilde{A}}} $'s $k$-th element stores the detected codeword numerical tag $n$. $k$ is the row numerical tag of $\tilde{\mathbf{X}}$. The operation of line 6 assigns all elements in a row into zero and thus guarantees each row is matched with only one codeword. It has to be clarified that the elements of restored $\tilde{\mathbf{X}}$ is the transmitted symbols at each channel uses, yet remained to be demodulated and the rows in the matrix contain both potential active devices and other redundant vectors. Algorithm 1 successfully resolves the permutation ambiguity during the dictionary learning by matching each row of restored information $\tilde{\mathbf{X}}$ or the column of the channel matrix with a possible codeword. Thus, the preliminarily desired information can be extracted, but the scalar ambiguity remains to be eliminated. 

\textit{3) Dictionary matrix refined by ECC and DL:} 
 Fig. 3 illustrates the overall structure of the receiver, including activity detection where the predefined sparse frame construction is utilized to extract information by Algorithm 1. The following will elaborate the rest procedures at length, especially on how to adopt ECC to eliminate the ambiguities encountered in DL algorithm. Generally, observations decomposition and approximation improvement are two majors procedures for DL. The latter can be achieved by conducting dictionary or coefficients matrix refinement strategies simultaneously or solely to achieve more accurate approximation, such as MOD and K-SVD. This reveals the spirit of DL is to form a desired linear combination by a set of basis vector. Yet, algorithm cannot guarantee fully correct sparse recovery and the later perfect refinement, especially at the presence of outer noisy deviations, which puts impetus on further procedures in this proposed scheme. 
 
 \begin{figure}[!t]
  \centering
  \includegraphics[width=3.4in]{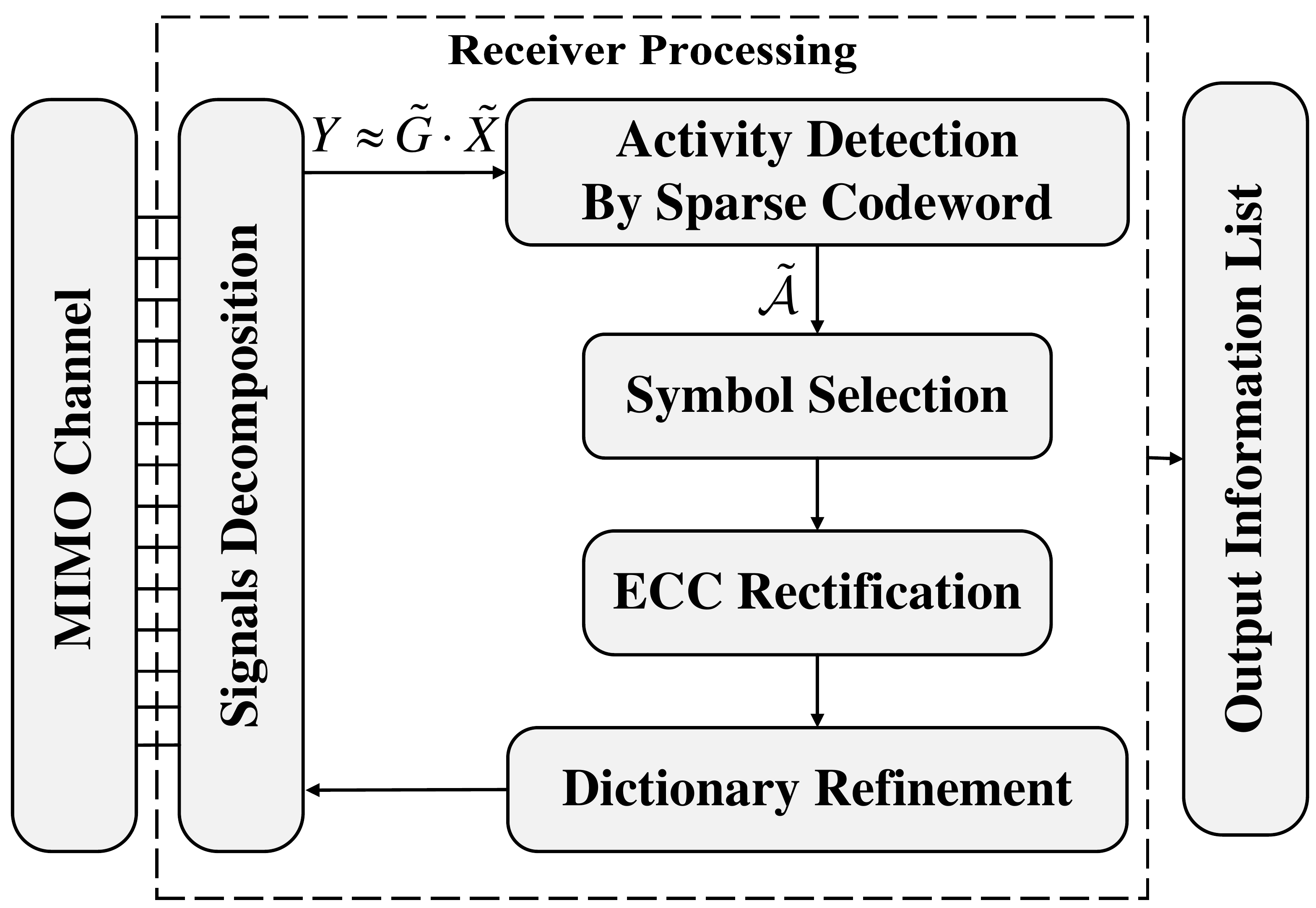}
  \caption{Illustration of the procedures at the receiver via DL and ECC methods.}
  \label{ReceiverProcedures}
\end{figure}

 After decomposition and active codeword detection, the dictionary refinement stage aims to conquer the scalar ambiguity and recover the desired information. Recalling that the recovered information row vectors $\tilde{\mathbf{x}}_{u}^{\text{T}}$ was matched with the most compatible codeword $\tilde{\mathbf{c}}_{u}^{\text{T}}$ by the inherent the sparsity pattern. The zero elements in codeword indicate that $\tilde{\mathbf{x}}_{u}^{\text{T}}$ should have the same zero elements at the corresponding location. This can be summarized as following:
  \begin{equation}
  \label{dictionary refinement1}
  \begin{split}
    &\mathop{\arg\min}\limits_{\tilde{\mathbf{G}}, \tilde{\mathbf{X}}} \left \| \mathbf{Y-\tilde{G} \tilde{X}}  \right \|_{2}^{2},\\ 
    &\text{s.t.} \ \left[\mathbf{x}_{u}^{\text{T}}\right]_{i}=0 ,\ if \ \left[\mathbf{c}_{u}\right]_{i}=0
    \end{split}
  \end{equation}
  
  It's natural that the approximation becomes more accurate when the dictionary and the coefficients matrices are restored more correctly. (\ref{dictionary refinement1}) can be deemed as first refinement on the information matrix. Besides, If the errors in $\tilde{\mathbf{X}}$ can be detected and corrected, rendering the matrix into a more desired information matrix $\hat{\mathbf{X}}$, i.e., $\hat{\mathbf{X}}$ resembles the preliminary $\mathbf{X}$ much more than $\tilde{\mathbf{X}}$. Then, conducting dictionary refinement based on $\hat{\mathbf{X}}$ would make $\tilde{\mathbf{G}}$ evolve towards the actual channel matrix. Thus, the approximation is made more accurate.

  \begin{algorithm}[H]
    \caption{Dictionary refined by ECC and MOD }\label{alg:alg2}
    \begin{algorithmic}[1]
    \STATE {\textsc{INPUT}} $\tilde{\mathbf{X}}, \ \tilde{\mathbf{G}}, \ \mathcal{\tilde{A}}, \ \mathbf{C}, \ \mathbf{H}$
    \STATE Define $\bm{\mho}=\mathbb{EX}\left [ \mathbf{c}_{u} \right ] $ as pattern extracting matrix
    \STATE \hspace{0.5cm}$\textbf{For} \ i=1:K_{a} \ \textbf{do}$
    \STATE \hspace{1cm}$k=\textbf{find}(\tilde{\mathcal{A}}_{i\text{-th}}\ne 0)$
    \STATE \hspace{1cm}$\bm{\mho}=\mathbb{EX}\left [ \mathbf{c}_{\left [ \mathcal{A} \right ]_{k} } \right ]$
    \STATE \hspace{1cm}$\left [ \hat{\mathbf{X}}  \right ]_{i,:} = \left [ \tilde{\mathbf{X}} \right ] _{k,: } \cdot \bm{\mho} $
    \STATE \hspace{1cm}$\left [ \hat{\mathbf{G}}  \right ]_{:,i} = \left [ \tilde{\mathbf{G}} \right ] _{:,k} $
    \STATE \hspace{0.5cm}\textbf{end for}
    \STATE \textbf{Demodulation} $\hat{\mathbf{X}}_{\text{llr}}=\mathbb{DM}[\hat{\mathbf{X}}]$
    \STATE \textbf{ECC-Decoder} $\hat{\mathbf{X}}_{\text{binary}}=\mathbb{LDPC}^{-1}[\hat{\mathbf{X}}_{\text{llr}},\mathbf{H}]$ 
    \STATE Encode $\hat{\mathbf{X}}_{\text{binary}}$ into $f(\hat{\mathbf{X}}_{\text{binary}})$, the state during transmission, to modify the channel matrix
    \STATE \textbf{Dictionary refinement} $\hat{\mathbf{G}}=\mathbb{MOD}[f(\hat{\mathbf{X}}_{\text{binary}}),\mathbf{Y}]$
    \STATE {\textsc{OUTPUT}} $\hat{\mathbf{G}}$
    \end{algorithmic}
    \label{alg2}
    \end{algorithm} 

  Assuming successful active codeword detection, Algorithm 2 elaborates the refinement made by ECC and dictionary learning. In line 2, $\bm{\mho}$ is a matrix generated by the pattern of nonezero elements in a vector, aiming to extract the nonezero elements out of the original vector. Its column number equals to the total amount of nonezero elements in the vector and if the $m$-th nonezero element is located at the $n$-th sequential order, then $\left[\mathbf{\bm{\mho}}\right]_{n,m}=1$, e.g., for a row vector $\mathbf{v}=\left [ 1,0,1,1 \right ]$, $\mathbf{\bm{\mho}}=\left [ 1,0,0,0;0,0,1,0;0,0,0,1 \right ]^T \in \left \{ 0,1 \right \} ^{4\times 3}$ and $\mathbf{v}_{\text{nonezero}}=\mathbf{v\cdot \bm{\mho}}=\left [ 1,1,1 \right ] $. Line 4 means find the $i$-th nonezero element in the set. Line 6-7 reduce the redundant rows (columns) in $\tilde{\mathbf{X}}$ ($\tilde{\mathbf{G}}$) and line 6 extracts potential symbols by the selected codeword. In line 9-11, $\hat{\mathbf{X}}$ is firstly demodulated from symbols to LLR (Log-likelihood Ratio)-expressed value. Then, $\hat{\mathbf{X}}$ becomes the input of ECC-decoder, where we use LDPC soft decoding with belief propagation algorithm and $\mathbf{H}$ is the check matrix. After the correction in ECC, to conduct dictionary refinement, every row of $\hat{\mathbf{X}}_{\text{binary}} \in \left \{ 0,1 \right \} ^{K_{a}\times B}$  is encoded into $f(\hat{\mathbf{X}}_{\text{binary}})\in\mathbb{C} ^{K_{a}\times L}$ because of the constraint subordinated to (\ref{DLtoCS}) and $f\left ( \cdot  \right ) $ aims at encoding the row vectors of the matrix. The final move in line 12 is to make dictionary refinement, MOD algorithm is adopted to make dictionary modification for its easy implementation and guaranteed convergence. 
  
  Note that the scalar ambiguity are implicitly solved during line 9-12, because after demodulation and ECC-decoding and encoding $f(\cdot)$, the scalar matrix $\mathbf{\Lambda}$ is no longer an arbitrary complex scalar matrix but just a matrix $ \mathbf{I}=diag\left ( \left [ a_1,a_2,\dots ,a_K \right ] \right ) $ whose element is none-complex, i.e., $\mathbf{DX}=\hat{\mathbf{D}}\mathbf{\Lambda}^{-1}\cdot   \mathbf{\Lambda}\hat{\mathbf{X}}$ where $\mathbf{\Lambda}$ used to be arbitrary and unpredictable during DL decomposition stage, whereas now its influence is negligible because its elements are either positive or negative constants. In terms of the amplitude of information, it barely cripples the mapping demodulation and ECC-decoding due to the unified scalar multiplication, i.e., the whole message vector multiplied with a scalar won't affect the information outcomes in the decoding stages, e.g., when QPSK is adopted, the demodulated symbols vector can be expressed in the form of approximate LLR counteracting the amplitude scalar multiplication and thus solve the scalar ambiguity. Speaking of impact on phase, the worst case is the phase reversal of the whole vector, which can be feasibly resolved using Differential encoding method, such as Manchester code.
  
\subsection{Collision Resolution}
It's possible that two or more devices select the same codeword, $\tilde{\mathbf{c}}$, when all users share a common codebook, which, as referred in Section III-A, may occur in a small probability in this scheme. When a codeword is selected by multiple devices to control the transmission pattern, subsequently, the resulted received signals at receiver can be expressed as
\begin{equation}
\label{COLLIDE}
\begin{split}
  \mathbf{Y}_{\text{collide}}
&=\sum_{i\in C}\mathbf{h}_{i}\mathbf{x}_{i}^{\text{T}}+\mathbf{N}   \\
             &=(\sum_{i\in C}\mathbf{h}_{i}\mathbf{\Lambda}_{i} )\tilde{\mathbf{c}}^{\text{T}}+\mathbf{N} 
\end{split}
\end{equation}
where $\tilde{\mathbf{c}}^{\text{T}}$ is the codeword in collision and $\mathbf{\Lambda}_{i}\tilde{\mathbf{c}}^{\text{T}}=\mathbf{x}_{i}^{\text{T}}$, $\mathbf{\Lambda}_{i}$ is a diagonal complex matrix, $\mathbf{x}_{i}^{\text{T}}\in\{0,1\}$ is the binary messages and $\mathbf{N}$ is the disturbances and $C$ is a set containing devices in collision. Under the proposed scheme, it means the constraints $\bm{\Omega}_{u}^{\text{T}}\cdot \mathbf{c}_{u}=S$ and $\bm{\Omega}_{u}^{\text{T}}\cdot \mathbf{c}_{\tilde{u}}\ll S$ in Section III-B are invalid, which results in failures of active user identification and the later stages. Collision may happen when one codeword is detected as active to multiple devices by Algorithm 1. This can be the result of overlapping in (\ref{COLLIDE}) or outer disturbances. And whether it's the cause of outer noises or there is indeed collision happening needs to be determined. Even though the chances of collision are trivial due to the abundant amount of potential codewords as referred in Section III-A, it can happen anyway and a collision resolution protocol aiming to prevent such thing is described in Algorithm 3. In line 4, $\bm{\mho}_{i}$ is the pattern extracting matrix defined in Algorithm 2 and is generated by codeword $\mathbf{c}_{i}$. Line 6 stands for the error detection procedure where the parity check is utilized for instance.

\begin{algorithm}[H]
  \caption{Collision resolution protocol }\label{alg:alg3}
  \begin{algorithmic}[1]
  \STATE When one codeword is detected as active to $d$ multiple device and $d > m$, collision may happen. $m$ is the tolerable codeword repetition number.
  \STATE \textbf{do} the following protocol
  \STATE \hspace{0.5cm}\textbf{Traverse} codewords to extract information
  \STATE \hspace{0.5cm}$\hat{\mathbf{x}}_{i}^{\text{T}}=\hat{\mathbf{x}}^{\text{T}}\cdot\bm{\mho}_{i}$
  \STATE \hspace{0.5cm}\textbf{Demodulation} $\hat{\mathbf{x}}^{\text{T}}_{i, \text{binary}}=\mathbb{DM}[\hat{\mathbf{x}}^{\text{T}}_{i}]$
  \STATE \hspace{0.5cm}\textbf{ECC-Error-Number} $e_{i}=\mathbb{PARITY}[\hat{\mathbf{x}}^{\text{T}}_{i, \text{binary}}]$
  \STATE \hspace{0.5cm}Select $c_{i}$ with \textbf{min} $e_{i}$ and \textbf{max} inner product $\hat{\mathbf{x}}^{\text{T}}_{i}\cdot  c_{i}$ 
  \STATE \textbf{end}
  \STATE {\textsc{UPDATE}} $\mathcal{A}$
  \end{algorithmic}
  \label{alg3}
  \end{algorithm} 

\subsection{Atom Number Optimization}
Recalling (\ref{CS1}) and (\ref{DL2}), the sparse approximation via dictionary learning generates coefficients or atoms, $\mathbf{\tilde{x}}$, from measurements $\mathbf{y}$, by a set of basis vectors in dictionary. Normally, by the nature of sparse approximation, the size of nonezero elements in $\mathbf{\tilde{x}}$ is much smaller compared with whole measurements and the demand for reliable recovery and acceptable computational complexity differs with the atom number, for instance, \cite{ref116} demonstrates theoretically and empirically that the OMP algorithm requires at least $\mathcal{O}(m\ln{d} )$ measurements of the signal and $\mathcal{O}(mNd )$ computational cost for a column vector $\mathbf{x}$ in $d$ dimension with $m$ none-zero atoms and a row-wise $N$ dimension dictionary matrix equaling the size of measurements. 

Thus, when adopting the dictionary learning algorithm in massive access, how to properly initialize a reasonable amount of atoms or the row-wise sparsity level of $\mathbf{x}$ needs to be contemplated. The below elaborates how to bridge the statistic feature of codebook and the sparsity level of the recovered information in order to achieve relatively low computational complexity and a reduced antenna size at the BS. In (\ref{DLtoCS}), $\mathbf{X}\in\mathbb{C}^{K_{tot}\times L}$ and $\mathbf{G}\in\mathbb{C}^{M\times K_{tot}}$, $M$ is the number of antennas at the BS, $L$ is the frame length. Only $K_{a} \ll K_{tot}$ rows denoting active devices at random permutations in $\mathbf{X}$ have none-zero entries, leading to a basic idea that the upper bound of the row-wise atom number $m_{max}=K_{a}$ by which the atom number of dictionary learning can be simply set. However, an upper bound barely reflects the row-wise sparsity level. To obtain the appropriate atom number, a new train of thought is provided. Provided that the common codebook is generated at solely random, specifically, it means every element in a single codeword has a $\gamma =S/L \ll 1$ likelihood to be none-zero and the distribution of elements in a column is also random and irrelevant to that of rows, which is also the case in $\mathbf{X}$ since the randomly active devices transmit the messages by the corresponding sparse codeword. Thus, overall speaking, the elements in $\mathbf{X}$ matrix are i.i.d., by which the average number of none-zero elements $m$ of column vectors is $(K_{a}\cdot S )/L=K_{a}\cdot \gamma \ll K_{a}$, where $K_{a}\cdot S$ is the total none-zero entries in the matrix and $\gamma$ is determined by the design of codebook.

In this regard, for example, when OMP is adopted, the computational complexity (CC) for a matrix is proportionate to $\mathcal{O}(mMK_{tot}L)=\mathcal{O}(\alpha mK_{a} )$, where $\alpha=(MK_{tot}L)/K_{a}$ is assumed to remain as a constant. If the atom number is set as the upper bound $m=K_{a}$ or $m=K_{a}\cdot \gamma$, then the CC is $\mathcal{O}(\alpha K_{a}^{2})$ and $\mathcal{O}(\beta K_{a})$ respectively, where $\beta=\alpha K_{a}\cdot \gamma$. Conspicuously, CC is reduced greatly away from Exponential Growth. Similarly, the least request on the size of measurements for robust recovery is reduced with fewer number of atoms. Numerical results in Section IV validates the conducive effect on the outcomes by utilizing the statistic feature of codebook to offer guidance on choosing the number of atoms. The above is illuminating for the extension to the scenario where the row of codebook follows various distributions and where the impact from row vectors on the statistic features of column vectors needs to be considered.
\subsection{Resolving The Unknown $K_{a}$}
The previous parts deem the number of active devices as a known setting which in practice often needs to be estimated. Since the massive MIMO is equipped at the BS, by the Law of Large Numbers, an upper bound of the potential devices can be obtained from the power estimation 
\begin{equation}
  \label{K_Estimation1}
  \hat{K}= \frac{1}{\check{\rho}  S}\left ( \frac{\left \| \mathbf{Y} \right \|_{F}^{2} }{M} -L \right  ) 
  \end{equation}
where $\check{\rho}$ is the least received symbol power for the receiver. A soluble method for remedy is to initialize the dictionary learning with an estimated upper bound active device number $\hat{K}$ estimated by (\ref{K_Estimation1}) from the lower symbol power bound value and then filter out the unqualified potential rows in the outcomes of dictionary learning stage by considering both the power of the atoms and the inner product value to determine the final estimation $K_{\text{est}}$. The Algorithm 4 elaborates the matrix trimming process with the initialized value $\hat{K}$ and the active codeword detection in Algorithm 1. 

\begin{algorithm}[H]
  \caption{Matrix trimmed by pattern and power }\label{alg:alg4}
  \begin{algorithmic}[1]
  \STATE {\textsc{INPUT}} $\hat{K}$, $\mathbf{Y}$, $\mathbf{ \tilde{X} }$, $\mathbf{ \tilde{G} }$,$\mathbf{\mathcal{\tilde{A}} }$ 
  \STATE Initialization with $\hat{K}$
  \STATE \textbf{while} $i\le \left \| \tilde{\mathcal{A}}   \right \|_{0} $ do
  \STATE $k=\textbf{find}(\tilde{\mathcal{A}}_{i\text{-th}}\ne 0)$
  \STATE $\mathbf{x}_{i}^{\text{T}}=\left [ \mathbf{ \tilde{X} }\right ]_{k,:}\cdot \mho_{i}$, $\mathbf{h}_{i}=\left [  \mathbf{ \tilde{G} } \right ]_{:,k}$
  \STATE \textbf{if} $\left \| \mathbf{x}_{i}^{\text{T}} \right \| _{2}^{2}\cdot c_{i}< threshold$ 
  \STATE \hspace{0.5cm}$ \hat{K}=\hat{K}-1 $
  \STATE \hspace{0.5cm}exclude $\mathbf{x}_{i}^{\text{T}}$ and $\mathbf{h}_{i}$ from $\mathbf{ \tilde{X} }$ and $\mathbf{ \tilde{G} }$
  \STATE \textbf{end if}
  \STATE \textbf{end while}
  \STATE $K_{\text{est}}=\hat{K}$
  \STATE {\textsc{OUTPUT}} $ K_{\text{est}} $
  \end{algorithmic}
  \label{alg4}
  \end{algorithm}  
  
\section{NUMERICAL RESULTS}

\begin{figure*}[!t]
  \centering
  \subfloat[]{\includegraphics[width=3in]{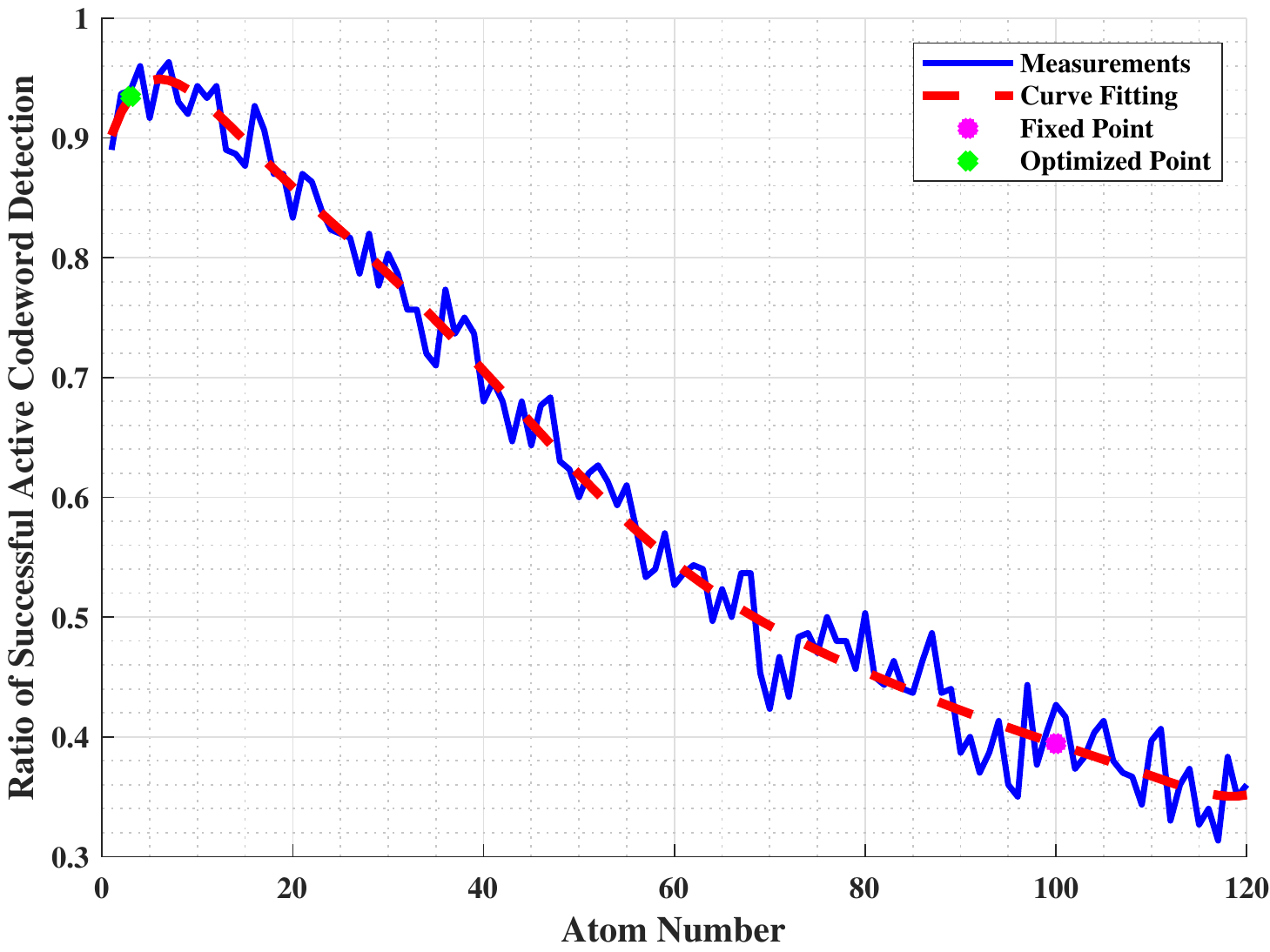}%
  \label{Atom Number 1}}
  \hfil
  \subfloat[]{\includegraphics[width=3in]{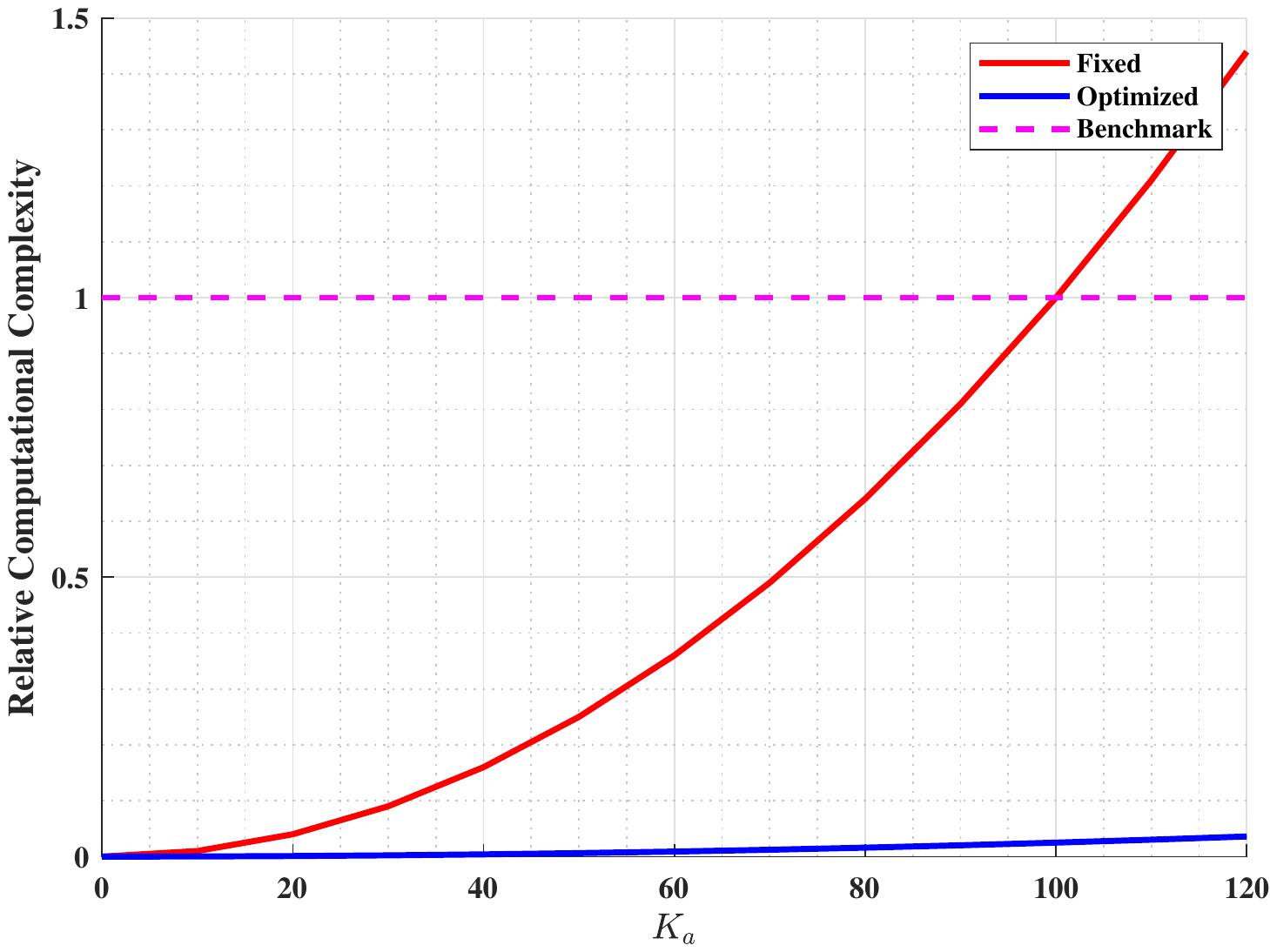}%
  \label{Atom Number 2}}
  \caption{The impact of atom number selection on activity detection and computational complexity. (a) Ratio of successful active codeword detection versus atom number with $K_{tot}$=1000, $K_{a}$=100, $M$=64, $L$=1600,  $S$=$L$/40, SNR=10dB. (b) Computational complexity versus atom number with $K_{tot}$=1000, $M$=64, $L$=1600,  $S$=$L$/40.}
  \label{Atom_Number_Selection}
  \end{figure*}

In this section, simulation results demonstrate the viability of the proposed scheme under unsourced massive access with a massive MIMO BS scenario. In the simulation, all devices share a randomly generated common codebook and encode binary messages by ECC method. Then, the ECC-coded messages are mapped into symbols via the quadrature phase shift keying (QPSK) and transmitted by the sparse frame pattern of the selected codewords. The $L$-length frame consists of $S$ none-zeros and $(L-S)$ silent compartments, whose permutations are determined by the codebook. Perfect synchronization between devices and receiver is assumed. No pilot preambles or pilot-like bits were utilized and the i.i.d. Rayleigh fading channel by drawing channel vector $\mathbf{h}_{k}$ from $\mathcal{CN}\left ( \mathrm {0},\mathbf {I} \right )$ was assumed to be constant during a frame time. The binary messages were LDPC-encoded at 0.5 code rate and decoded by BP algorithm\cite{ref99} in ECC procedure and the dictionary learning stage was carried out by OMP and MOD algorithms. Unless otherwise noted, no collision takes place between devices and the number of active devices $K_{a}$ is assumed to be known and the power of symbols is identical.

\subsection{Active Codeword Identification}
The system performance is denoted in terms of per-user probabilities of error (PUPEs) \cite{ref7}, including \textit{per-user probability of misdetection} $p_{\text{md}}$ and \textit{per-user probability of false alarm} $p_{\text{fa}}$. The former is defined as 
\begin{equation}
  \label{Denoter1}
  p_{\text{md} }= \frac{\mathbb{E} \left [n_{\text{md} }\right ] }{K_{a}}  
\end{equation}
where the expectation is taken due to the randomly generated codebook, the fading, and the noise and $n_{\text{md}}$ stands for the number of the transmitted, yet eventually omitted messages. Empirically, since we assume all devices transmit same length messages, $n_{\text{md}}$ equals the amount of omitted active codewords. The latter is defined as
\begin{equation}
  \label{Denoter2}
  p_{\text{fa} }=\mathbb{E} \left [  \frac{n_{\text{fa} }}{\left | \mathcal{L} \right |  }\right ]  
\end{equation}
which is correlated to the estimated number of active codewords, denoted by
\begin{equation}
  \label{Denoter3}
  \left | \mathcal{L} \right |=n_{\text{fa} }+K_{a}-n_{\text{md} }
\end{equation}
$n_{\text{fa} }$ is the number of false alarms, i.e., the detected messages were actually never transmitted by any devices. Noting that when the number of active devices $K_{a}$ is known at the BS, $\left | \mathcal{L} \right |=K_{a}$ and consequently the error probability is equivalent $p_{\text{fa} }=p_{\text{md} }$, denoted as $p_{e}$ if the active number is a known. The zeros in codewords mean the silent slots where no signals are transmitted, but $\bar{\rho} $ is the averaged received power per symbol of each device and noise variance $\sigma^2$ is set to 1 without loss of generality. To this end, the energy-per-bit is defined as
\begin{equation}
  \label{energy-per-bit}
    E_{b}/ n_{0}\triangleq \frac{\bar{\rho} \cdot S}{2L\cdot \sigma^2}    
\end{equation}

\subsubsection{Choices of Atom Number}
%\begin{figure}[!t]
%\centering
%\includegraphics[width=3.2in]{AtomNumber1.eps}
%\caption{Ratio of successful active codeword detection versus atom number with $K_{tot}$=1000, $K_{a}$=100, $M$=64, $L$=1600,  $S$=$L$/40, SNR=10dB}
%\label{Atom Number 1}
%\end{figure}
Fig. 4(a) shows the ratio of successful active codeword detection under different choices of atom number when adopting dictionary learning algorithm. $K_{a}$=100 active devices out of $K_{tot}$=1000 potential single-antenna devices transmit $L$-length frames embedded with $S$ symbols with SNR=10dB to the BS with $M$=64 antennas. No preambles are assigned. The doted line is fitted from discrete measurements to illustrate the viability of the selection of atom number based on the codebook statistic features. The fixed point in the diagram is the upper bound of the atom number. The results tell the fact that too large the number of atom number selected may cause redundancy in dictionary learning procedure, depleting the robustness in identifying active codeword. Fig. 4(b) compares the computational complexity between the optimized selection and the upper bound atom number in decomposing the received signal $\mathbf{Y}$ using OMP algorithm. The relative value of complexity is unified by the value of red curve at $K_{a}=100$. The red curve depicts the computational complexity with atom number equal to $K_{a}$, the upper bound. And the blue curve depicts the counterpart with optimized atom number. The surging speed is comparably slower with optimized atom number choices. 
%\begin{figure}[!t]
%\centering
%\includegraphics[width=3.4in]{AtomNumber2.eps}
%\caption{Computational complexity versus atom number with $K_{tot}$=1000, $M$=64, $L$=1600,  $S$=$L$/40}
%\label{Atom Number 2}
%\end{figure}

\begin{figure*}[!t]
  \centering
  \subfloat[]{\includegraphics[width=3in]{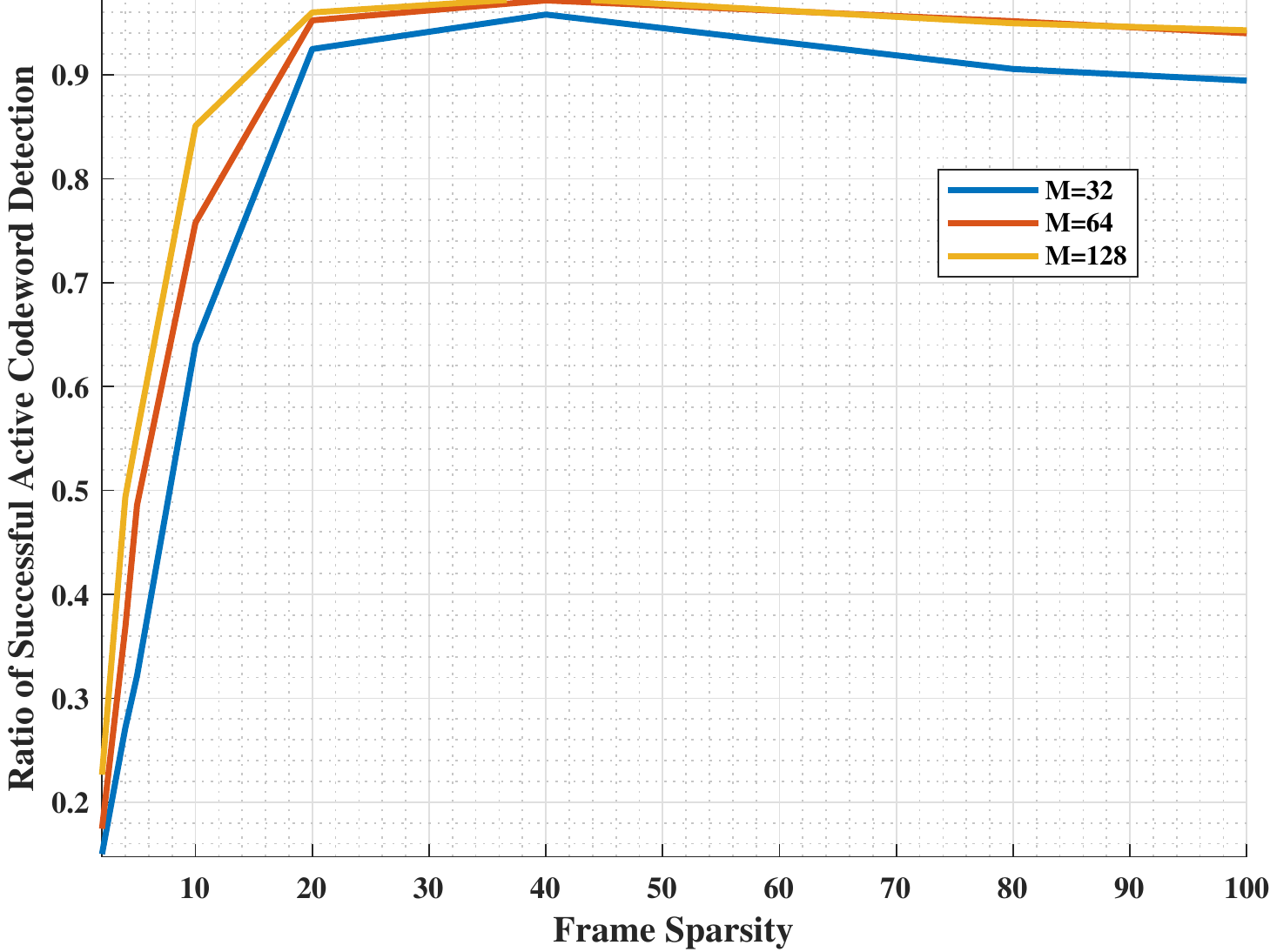}%
  \label{fig_first_case}}
  \hfil
  \subfloat[]{\includegraphics[width=3in]{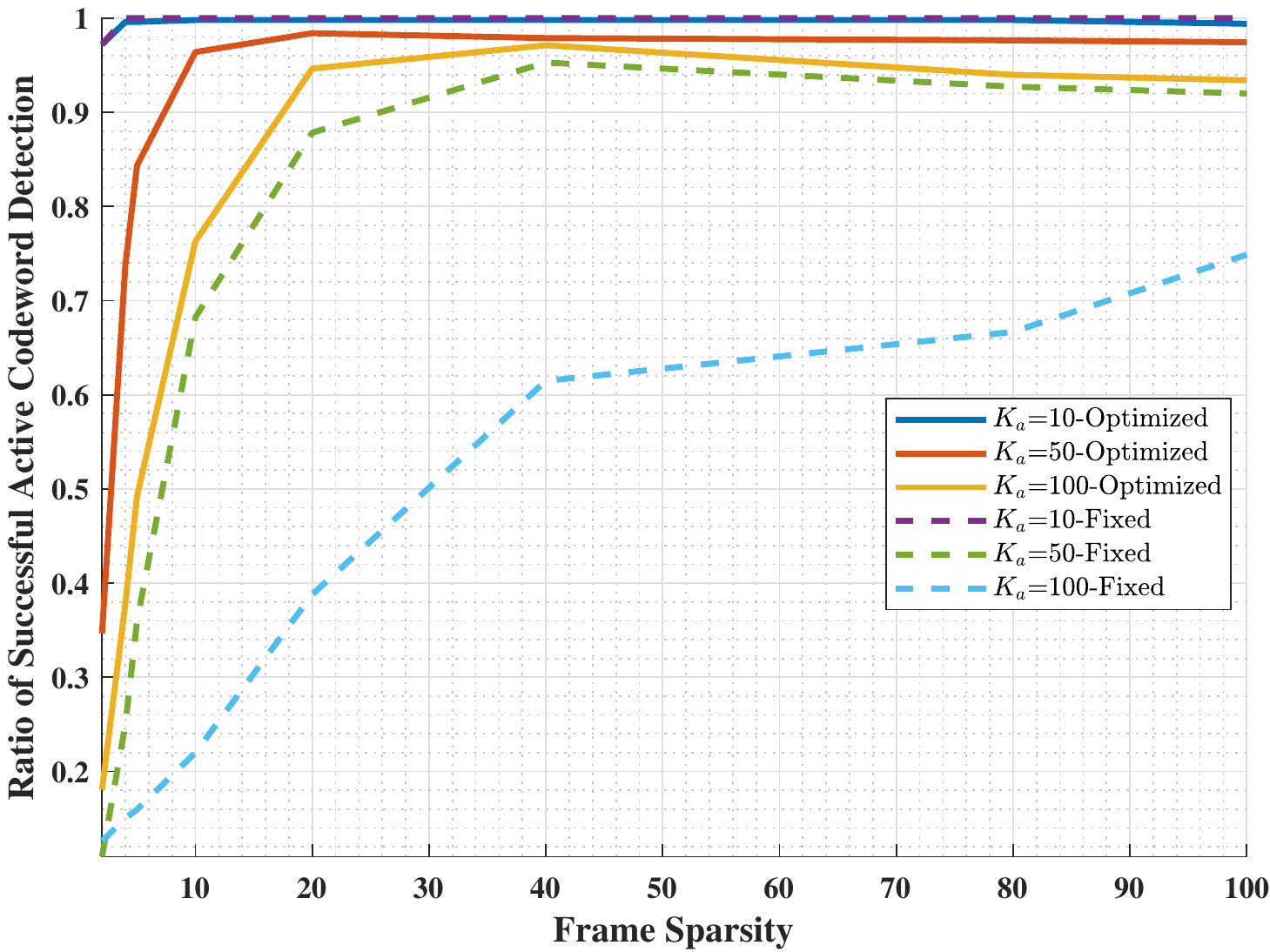}%
  \label{fig_second_case}}
  \caption{ (a) Ratio of successful active codeword detection versus frame sparsity with $K_{tot}$=1000, $K_{a}$=100, $L$=1600, SNR=10dB under different antennas number at BS . (b) Ratio of successful active codeword detection versus frame sparsity with $K_{tot}$=1000, $M$=64, $L$=1600, SNR=10dB under different atom number selection.}
  \label{fig_sim}
  \end{figure*}
As the compressed sensing theory implies, the least required measurements size for sparse recovery increases with more atom number selected. In this paper, the measurements size is equivalent to the number of antennas at the BS. Fig. 5(a) shows how the identification performance alters under different antenna number equipped at the BS with frame sparsity ($L/S$)  changing. A modest frame sparsity guarantees favorable identification outcomes after $L/S=20$ for less redundancy and a relatively sufficient information provided. All curves hold a decreasing tendency after $L/S=40$ due to increasingly insufficient information. However, similar performance after frame sparsity 20 when the antenna size is chosen as $M=64$ and $M=128$ indicate that fewer antennas are needed to support certain size users by flexibly choosing atom number in the algorithm, which above all offers a guidance on the antenna options concerning different size of potential users.  
%\begin{figure}[!t]
%\centering
%\includegraphics[width=3.2in]{AtomNumber3.eps}
%\caption{Ratio of successful active codeword detection versus frame sparsity with $K_{tot}$=1000, $K_{a}$=100, $L$=1600, SNR=10dB}
%\label{Atom Number 3}
%\end{figure}
 Fig. 5(b) illustrates how frame sparsity affects successful identification ratio under different number of active devices and the choices of atom number. The doted lines set atom number by the upper bound and the solid lines otherwise by the statistically optimized atom number. When the number of active devices is small, both atom number selections identify all the active codeword without mistakes, whereas as $K_{a}$ increases larger, the upper bound selection performances reveal themselves much inferior. The above validates the viability of optimized atom number choices, the following simulations are conducted on this basis.
%\begin{figure}[!t]
%\centering
%\includegraphics[width=3.2in]{AtomNumber4.eps}
%\caption{Ratio of successful active codeword detection versus frame sparsity with $K_{tot}$=1000, $M$=64, $L$=1600, SNR=10dB}
%\label{Atom Number 4}
%\end{figure}

\subsubsection{Per-User Probabilities Of Errors}

\begin{figure}[!t]
\centering
\includegraphics[width=3in]{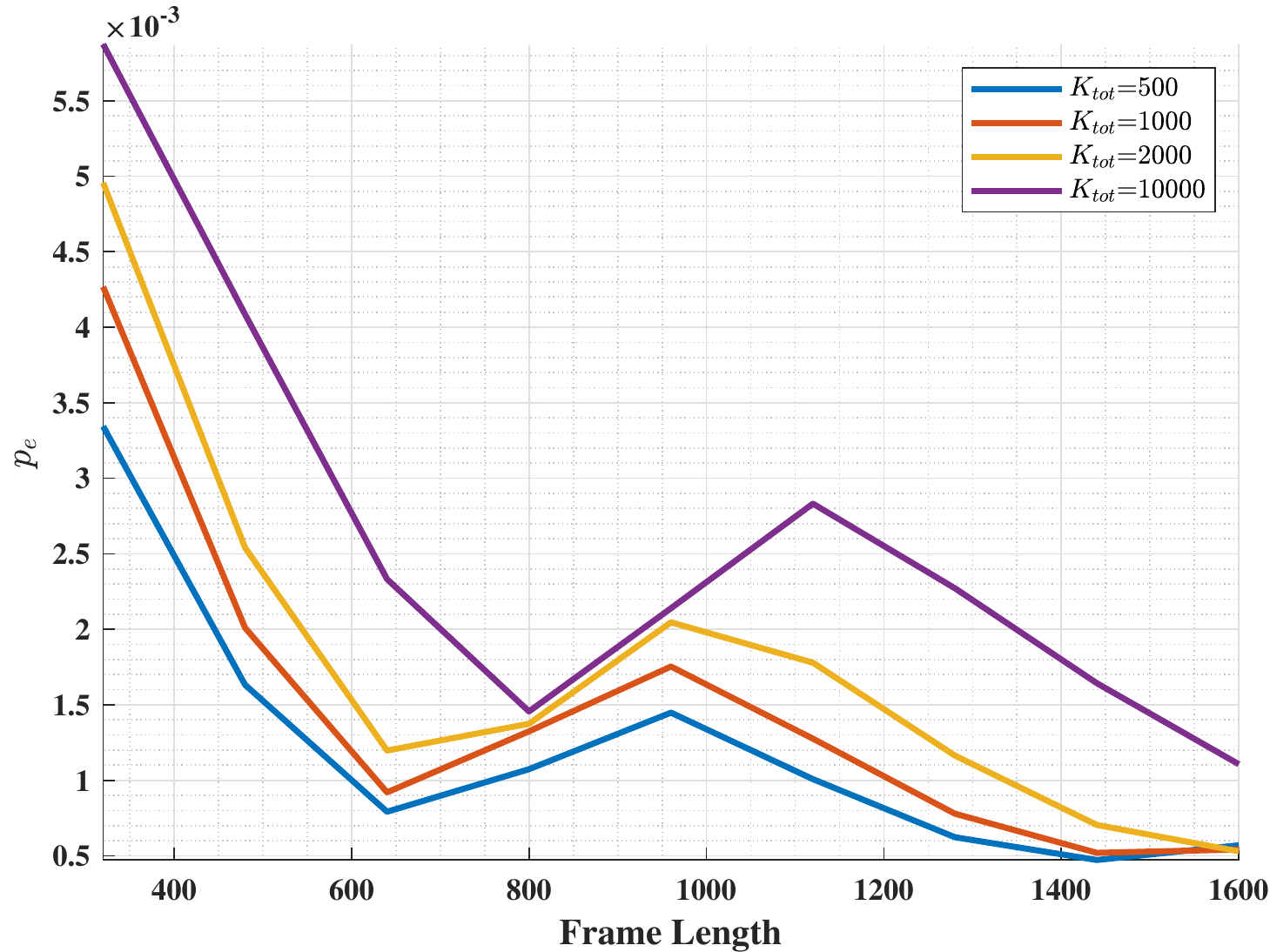}
\caption{PUPEs versus frame length with $K_{a}$=100, $M$=64, $L/S$=40, SNR=10dB}
\label{Atom Number 5}
\end{figure}

\begin{figure}[!t]
  \centering
  \includegraphics[width=3in]{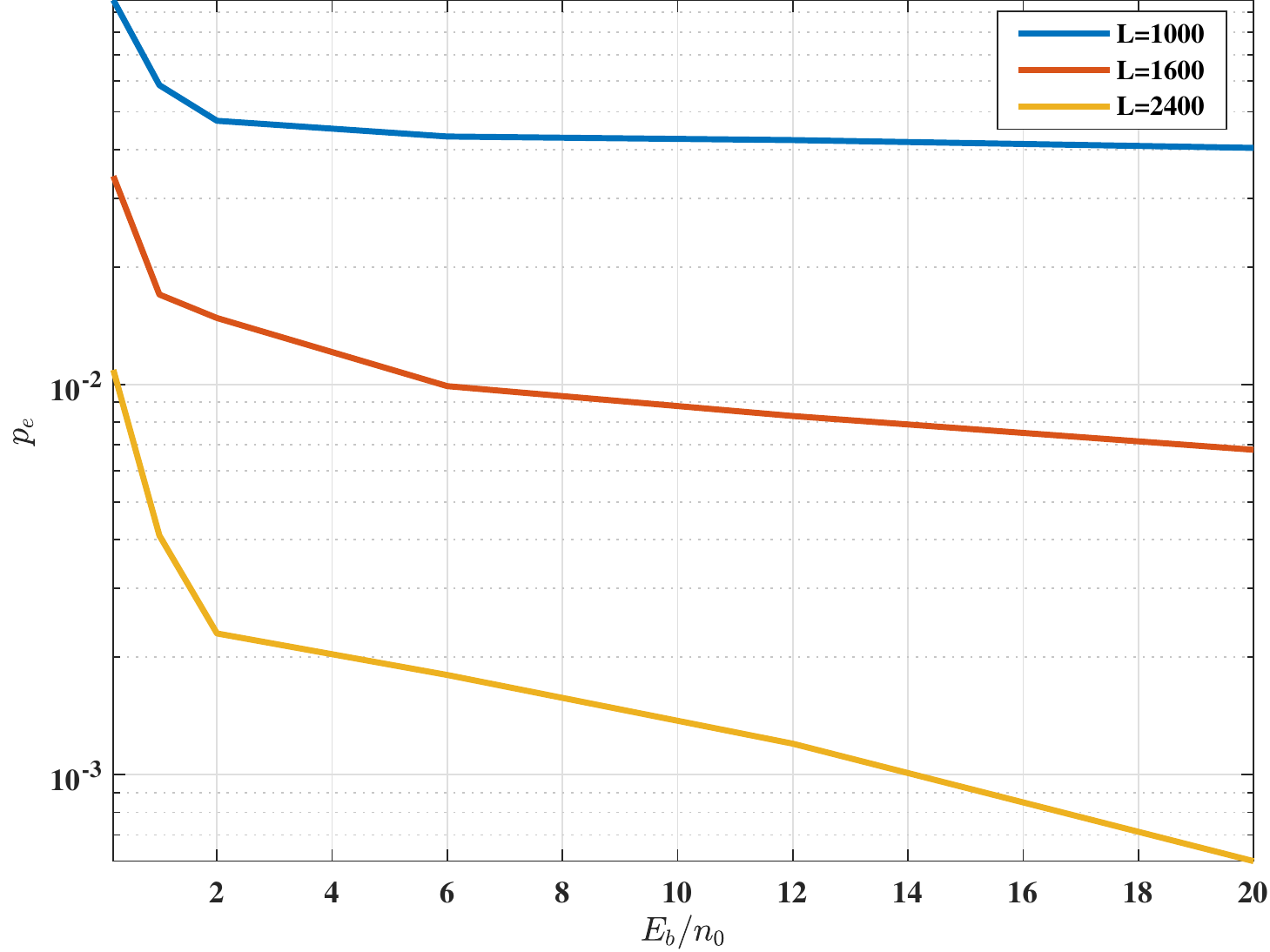}
  \caption{PUPEs versus $E_{b}/ n_{0}$ with $K_{tot}$=1000 ,$K_{a}$=100, $M$=64, $L/S$=100}
  \label{PUPEs1}
  \end{figure}

In Fig. 6, under different setting of potential devices $K_{tot}$ and a fixed amount of active devices $K_{a}$=100, the simulation shows the change of per-user detection errors $p_{e}$ while adopting longer frame length with fixed frame sparsity $L/S$=40. The curve may fluctuate for a while but eventually decrease to a favorable level, given the fact that as the frame length becomes longer, combined with a modest selection of atom number in algorithm, the increased information promotes the identification precision of active codeword. It's notable that the $p_{e}$ will drop with longer frame no matter what the size of $K_{tot}$ is, i.e., the degradation in the successful codeword detection is insignificant with the size of codebook, which is conducive to the massive access scenario because the overall size of active user $K_{a}$ is often considerably small in comparison with codebook size. The combination with Fig. 5 and Fig. 6 offers flexibility on choosing suitable parameters to control trade off during practical applications, since the frame sparsity directly determines how many symbols there are. Fig. 7 illustrates the correlation between $p_{e}$ and energy-per bit under different frame length settings with fixed frame sparsity $L/S$=100. The degradation of PUPE by increased $E_{b}/ n_{0}$ can be observed with any frame length and augmenting the information volume with a fixed sparsity is conducive to the active codeword detection. Combined with observations with fixed frame length and various sparsity in Fig. 5, frame length also takes crucial part in detection performance improvement. As explained in \cite{ref8}, the frame length can be assumed up to 120,000. In occasions where the total frame length or frame sparsity is limited, flexible setting on modestly low frame sparsity or relatively longer frame length can increase the information volume while maintaining favorable detection outcomes. 

\subsubsection{Information Restoration}
In codeword detection procedure, the decomposition generates matrix $\tilde{\mathbf{X}}$ carrying information to be restored. Via correction from ECC decoder and further decomposition with new matrix $\tilde{\mathbf{G}}$ from dictionary refinement, the information is restored. Noting that only with the right codeword detection can the information embedded in the rows be restored, thus the following results will be the estimation from the correctly detected devices and the stop criterion is denoted by the total number of parity check, i.e., when the overall number of parity check error retains to certain level, the information restoration stage stops. 

\begin{figure}[!t]
  \centering
  \includegraphics[width=3in]{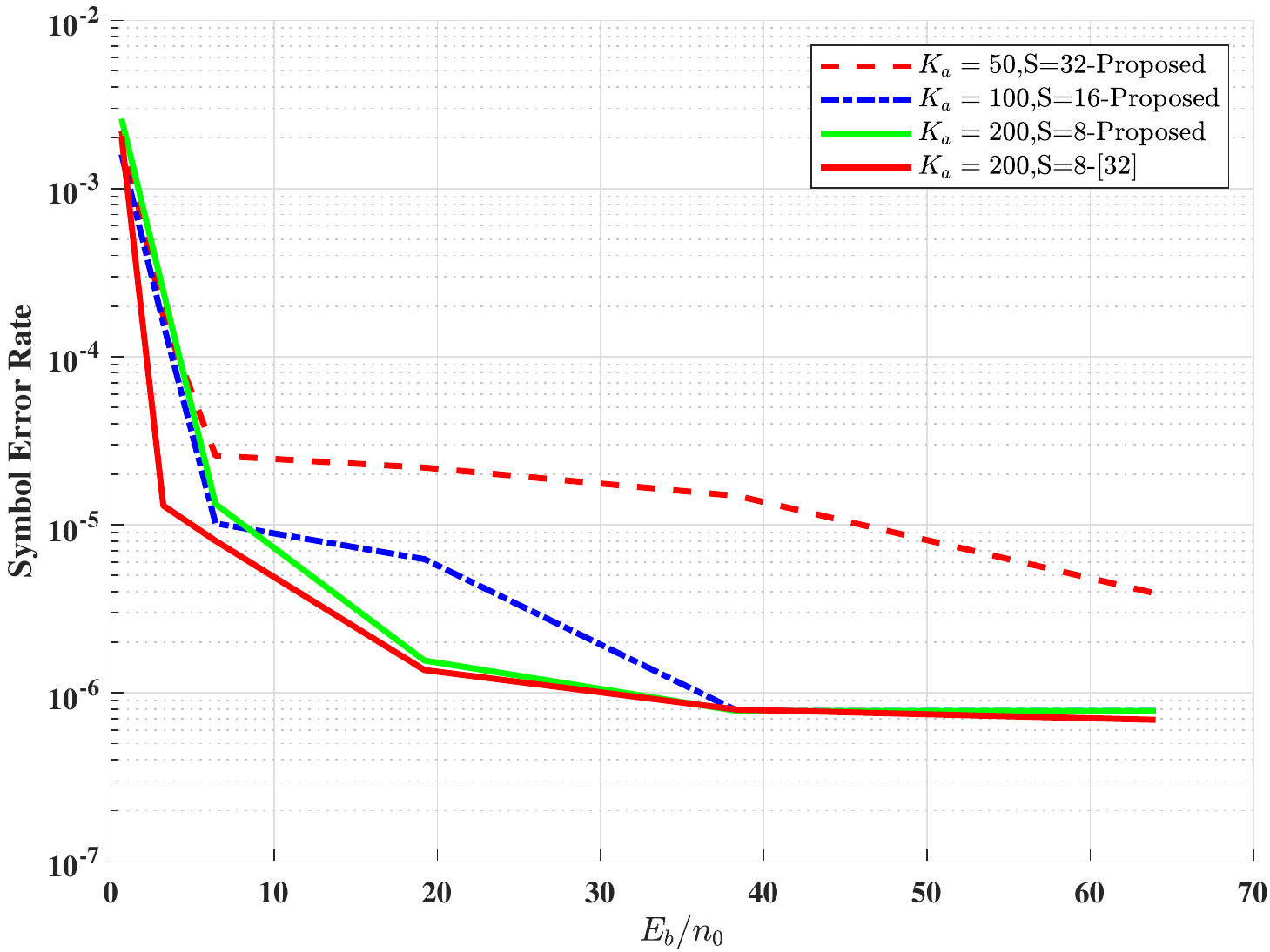}
  \caption{Symbol error rate versus $E_{b}/ n_{0}$ with $K_{tot}$=1000, $M$=64 under different $K_{a}$ and symbol number $S$.}
  \label{DataRestoration}
  \end{figure}

  \begin{figure}[!t]
    \centering
    \includegraphics[width=3in]{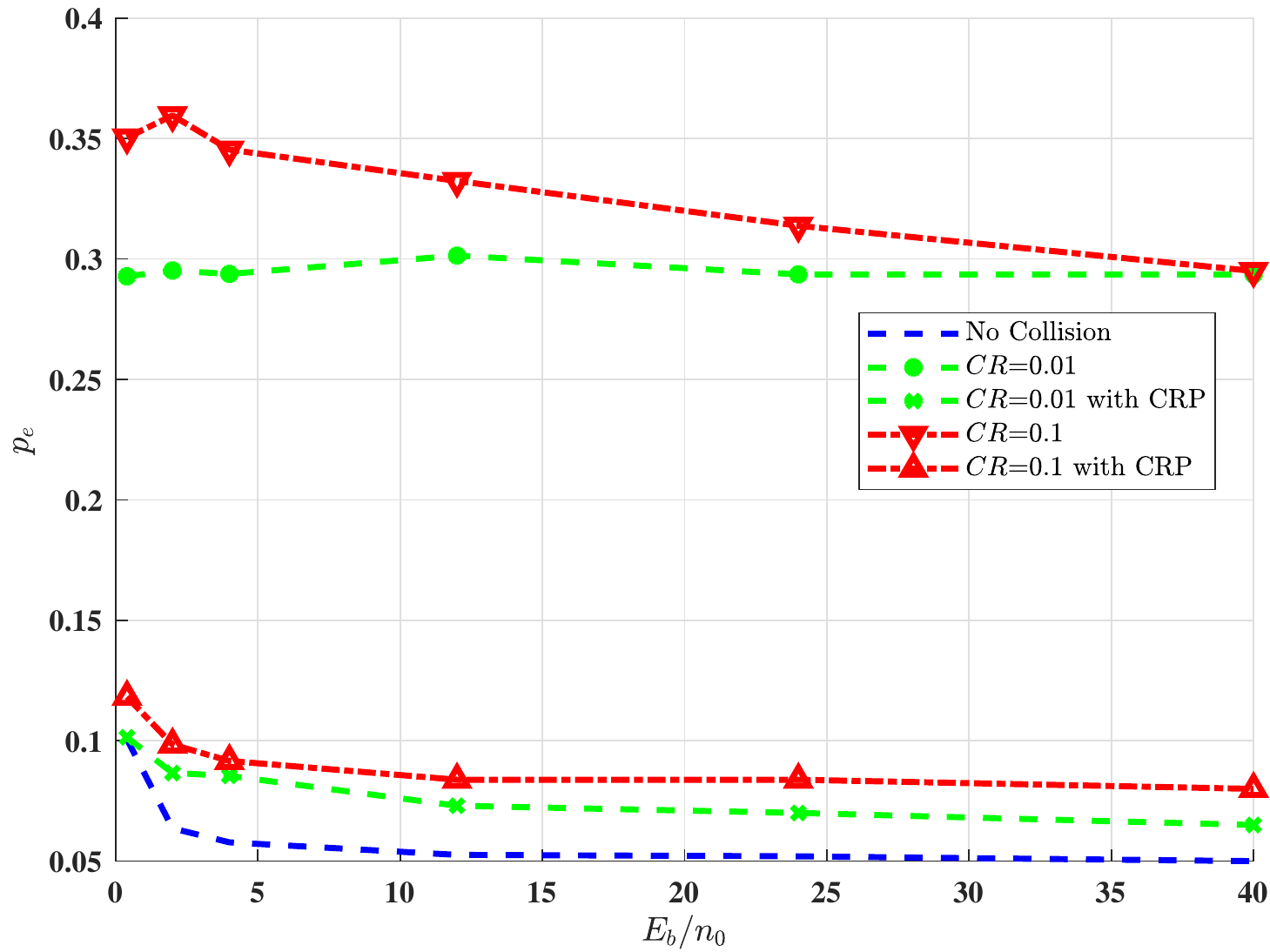}
    \caption{Comparison of $p_{e}$ between different collision ratio($CR$) with or without collision resolution protocol(CRP) and no collision circumstance versus $E_{b}/ n_{0}$ when $K_{tot}=1000$, $K_{a}=100$, $M=64$, $L=1000$, $S=20$, $m=2$.}
    \label{CollisionResolution}
    \end{figure}

The symbol error rate(SER) in Fig. 8 shows the information restoration under different amount of active devices and the transmitted symbols with an identical bits flow in total. Meanwhile, comparison is made between the proposed scheme and the scheme in \cite{ref122} whose frame structure incorporates unique user IDs in which several symbols act like pilots and are assumed to be priorly known information at the BS, and the receiver is also designed in the spirit of dictionary learning. The scheme with unique ID has a faster SER drop during a relative low $E_{b}/ n_{0}$ due to the blessing of the pilot-like bits and unique ID and shows similar performances with proposed scheme after $E_{b}/ n_{0}=18$. However, it's intolerable to restore all the priors of at the BS when the amount of potential devices is huge. A decreasing SER can be observed with larger $E_{b}/ n_{0}$ and when the number of active user becomes larger, shorter bit package outcomes favorable SER which fits the short package feature in mMTC. As the length of bits to be transmitted becomes shorter, the required energy-per-bit to reach certain SER level becomes smaller. It shows the advantage of small-packet by large sparse frame which decreases interferences among active devices greatly. The fewer symbols sent from the massive crowd, the easier the co-interference can be sort out.

\subsection{Collision Resolution Protocol}

  \begin{figure*}[!t]
    \centering
    \subfloat[]{\includegraphics[width=3in]{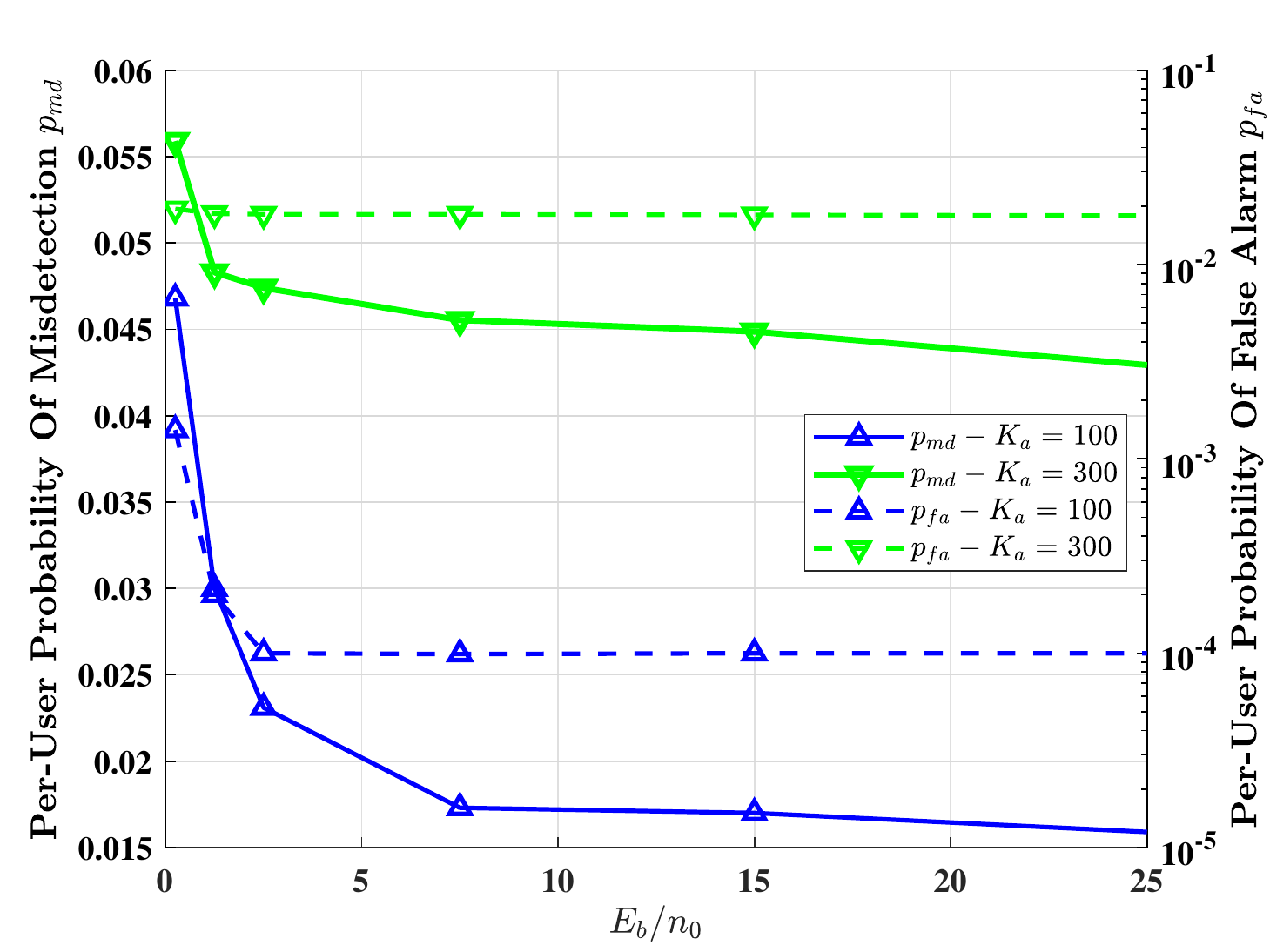}%
    \label{KaEstimation1}}
    \hfil
    \subfloat[]{\includegraphics[width=3in]{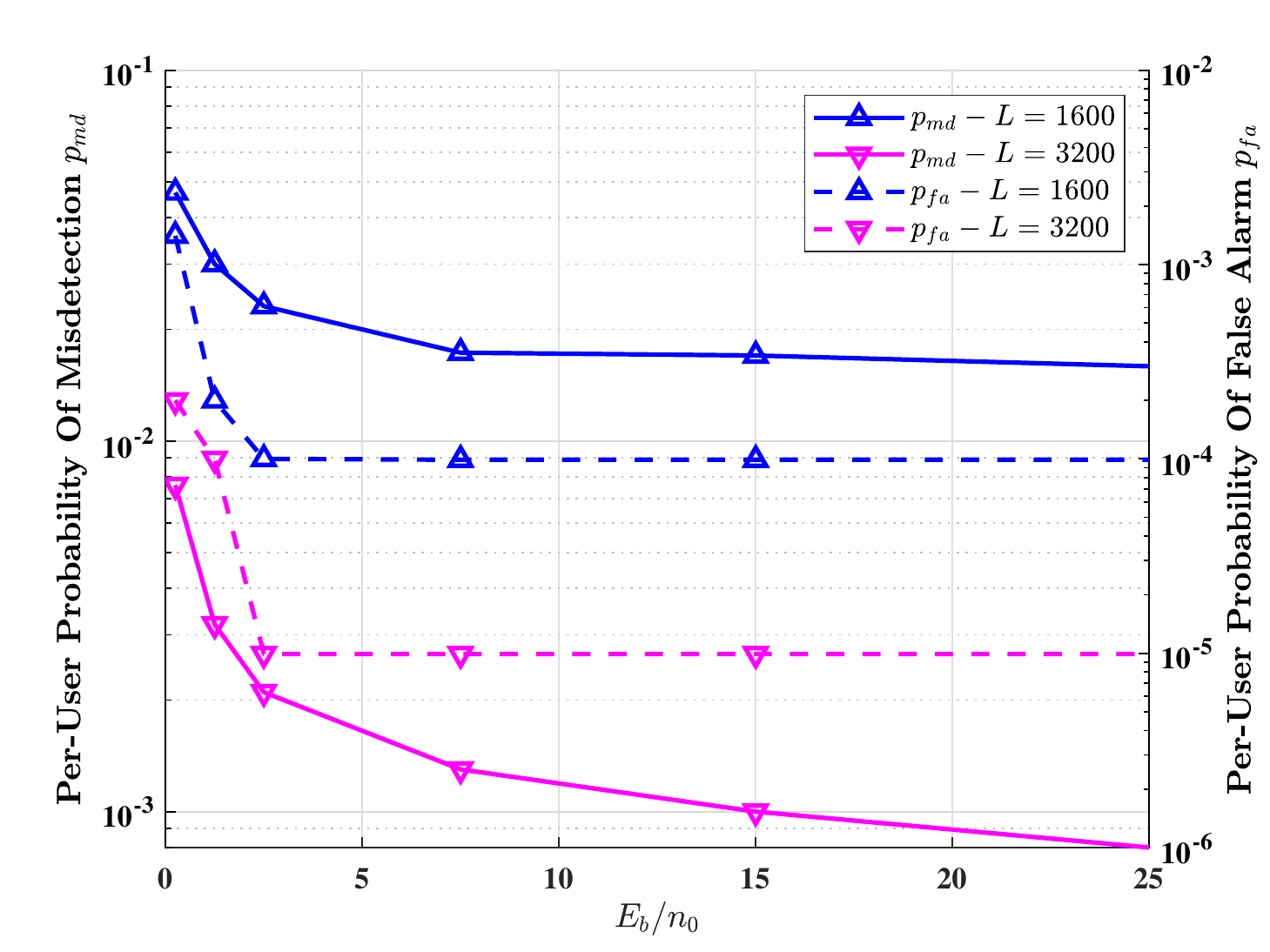}%
    \label{KaEstimation2}}
    \caption{Per-user probability of misdetection $p_{\text{md}}$ and per-user probability of false alarm $p_{\text{fa}}$ under $K_{a}$ estimation. (a) $p_{\text{md}}$ and $p_{\text{fa}}$ versus $E_{b}/n_{0}$ under active number estimation with actual active devices number of 100 and 300 when $K_{tot}=1000$, $M=64$, $L=1600$, $S=20$. (b) $p_{\text{md}}$ and $p_{\text{fa}}$ versus $E_{b}/ n_{0}$ under active number estimation with different frame length $L=1600$ and $L=3200$ when $K_{tot}=1000$, $K_{a}=100$, $M=64$, $L/S=80$.}
    \label{AKaEstimation}
    \end{figure*}

This part demonstrates the viability of the proposed collision resolution whose center idea is to diminish the false codeword detection caused by inaccurate decomposition in the existence of outer disturbances and to identify the collided codeword to continue the following procedures as much as possible by taking both the parity check error number and the value of inner product into consideration. Besides, since the collision likelihood is exceptionally trivial, the following renders the possibility of collision as a small ratio constant, and denotes the value of collision ratio by $CR$ and the abbreviation of collision resolution protocol as CRP. Meanwhile, the maximum number of codeword selection repetition is assumed to be a fixed constant $m$, i.e., every single codeword can only be chosen no more than $m$ times before transmission, which is a reasonable assumption due to the extremely low collision possibility. 

Fig. 9 illustrates the effectiveness of the proposed collision resolution protocol where the upper bound of codeword repetition $m=2$ and the $CR$ is set as 0.01 and 0.1 respectively, under which the protocol are adopted. After the adoption of the protocol, the $p_{e}$ in active codeword detection approaches the circumstance where no collision is assumed. Besides, the gap between the curve of CRP adopted and the curve with no collision gets bigger with larger $CR$, yet, usually, the possibility of collision is much lower.

\subsection{Estimation Of The Unknown $K_{a}$}

The above makes assumption that the amount of the active devices is a known constant which in actual practice is barely the case. By considering the impact from the atoms' power and the value of the inner product, the redundant rows in the estimation from (\ref{K_Estimation1}) are eliminated to improve the estimation accuracy of $K_{\text{est}}$ making preparations for the following procedures. When the number of active devices is an estimated value, the activity detection is divided into two performance indicators, per-user probability of misdetection $p_{\text{md}}$ and per-user probability of false alarm $p_{\text{fa}}$. 

In Fig. 10, the left vertical axis represents $p_{\text{md}}$ by solid curve and the right vertical axis represents $p_{\text{fa}}$ by doted lines. Fig. 10(a) illustrates the conducive impact from $E_{b} / n_{0}$ to PUPEs under different actual number of active devices and thus demonstrates the effectiveness of the proposed active number $K_{\text{est}}$ estimation in Algorithm 4. The performances under $K_{a}=300$ are inferior to their counterparts under $K_{a}=100$. However, due to the sporadic traffic in URA, the activity ratio normally is much less than 0.3. Fig. 10(b) extends the frame length to twice as much as that in Fig. 10(a) with $K_{a}=100$ and contains the same sparsity level. A comparable increase in estimation accuracy of devices number can be observed, owing to the growth of the information provided with a desirable sparsity level. These outcomes comply with Fig. 7 and validate the viability of the proposed estimation method.

%\begin{figure}[!t]
 % \centering
  %\includegraphics[width=3.2in]{KaEstimation1.eps}
  %\caption{$p_{\text{md}}$ and $p_{\text{fa}}$ versus $E_{b}/n_{0}$ under active number estimation with actual active devices number of 100 and 300 when $K_{tot}=1000$, $M=64$, $L=1600$, $S=20$.}
  %\label{KaEstimation1}
  %\end{figure}

  %\begin{figure}[!t]
    %\centering
    %\includegraphics[width=3.2in]{KaEstimation2.eps}
    %\caption{$p_{\text{md}}$ and $p_{\text{fa}}$ versus $E_{b}/ n_{0}$ under active number estimation with different frame length $L=1600$ and $L=3200$ when $K_{tot}=1000$, $K_{a}=100$, $M=64$, $L/S=80$.}
    %\label{KaEstimation2}
    %\end{figure}

\section{Conclusion}
In this paper, a DL and ECC-based unsourced random access scheme is proposed as a potential solution for a URA MIMO scenario. The scheme incorporates active codeword detection, information restoration, collision resolution and active device number estimation. The foundation is built on common codebook controlling sparsity pattern of frame and the utilization of DL with ECC. No pilot signal overhead is an appealing feature inaccurate. The numerical results validate the effectiveness of the scheme and illustrate the per-user possibility of errors (PUPEs) with certain vital parameters such as atom numbers, frame length and sparsity and energy-per-bit. Due to the difficulty in the analysis of DL, the performance of the proposed scheme was only numerically analyzed. Theoretical analysis would be crucial to the trade-offs between different parameters for system optimization. Besides, sparse common codebook design and research on asynchronous transmission scenario would be an important research branches in terms of practical URA system design.



\newpage

\vfill

\end{document}